\documentclass[a4paper,9pt]{article}    

\usepackage{latexsym}
\usepackage[english]{babel}  

\usepackage[dvips]{graphicx}         
\usepackage{subfigure}
\usepackage{longtable}     
\usepackage{amssymb}
\usepackage{epsfig,amsfonts}
\usepackage{color}
\usepackage{amsmath}                 
\usepackage{verbatim}                
\usepackage{syntonly} 
\usepackage{rotating}                
\usepackage{latexsym}
\usepackage{floatflt} 
\usepackage{eucal}
\usepackage{eufrak} 
\usepackage{graphics}
\usepackage{graphicx}
\usepackage{subfigure}
\usepackage{geometry}

\title{\bf{Identification of DNA-binding protein target sequences by physical effective energy functions. Free energy analysis of $\lambda$ repressor-DNA complexes}}

\begin{document}

\maketitle

\author{\bf{Elisabetta Moroni$^{a,b}$ \footnote{Corresponding author. \\ {\it Email addresses} moroni@to.infn.it (Elisabetta Moroni), caselle@to.infn.it (Michele Caselle), ffogolari@mail.dstb.uniud.it (Federico Fogolari)  }, Michele Caselle$^a$ and Federico Fogolari$^c$}}

\vspace*{1.1cm}

\noindent
$^a$ {\small Dipartimento di Fisica Teorica, Universit\`a di Torino and INFN, Via P. Giuria 1, 10125 Torino, Italy \\}
$^b$ {\small Dipartimento di Fisica G. Occhialini, Universit\`a di Milano-Bicocca and INFN, Piazza delle Scienze 3, 20156 Milano, Italy\\} 
$^c$ {\small Dipartimento di Scienze e Tecnologie Biomediche, Universit\`a di Udine, P.le Kolbe 4, 33100 Udine, Italy \\}

\vspace*{1.1cm}

\begin{center}
\section*{Abstract}
\end{center}         

Specific binding of proteins to DNA is one of the most common ways gene expression is controlled. Although general rules for the DNA-protein recognition can be derived, the ambiguous and complex nature of this mechanism precludes a simple recognition code, therefore the prediction of DNA target sequences is not straightforward.
DNA-protein interactions can be studied using computational methods which can complement the current experimental methods and offer some advantages.
In the present work we use physical effective potentials to evaluate the DNA-protein binding affinities for the $\lambda$ repressor-DNA complex for which structural and thermodynamic experimental data are available. 
The binding free energy of two molecules can be expressed as the sum of an intermolecular energy (evaluated using a molecular mechanics forcefield), a solvation free energy term and an entropic term. Different solvation models are used including distance dependent dielectric constants, solvent accessible surface tension models and the Generalized Born model.
The effect of conformational sampling by Molecular Dynamics simulations on the computed binding energy is assessed; results show that this effect is in general negative and the reproducibility of the experimental values decreases with the increase of simulation time considered.\\
The free energy of binding for non-specific complexes, estimated using the best energetic model, agrees with earlier theoretical suggestions. As a results of these analyses, we propose a protocol for the prediction of DNA-binding target sequences.
The possibility of searching regulatory elements within the bacteriophage $\lambda$ genome using this protocol is explored. Our analysis shows good prediction capabilities, even in absence of any thermodynamic data and information on the naturally recognized sequence.\\
This study supports the conclusion that physics-based methods can offer a completely complementary methodology to sequence-based methods for the identification of DNA-binding protein target sequences. 

\newpage

\section*{Introduction}
Protein-DNA recognition plays an essential role in the regulation of gene 
expression. Although a significant number of structures of DNA binding 
proteins have been solved in complex with their DNA binding sites increasing 
our understanding of recognition principles, most of the questions remain unanswered. 
Several studies showed that protein-DNA recognition could not be 
explained by a simple one-to-one correspondence between amino acids and bases \cite{Matthews:1988,Pabo:1984,Pabo:1992}, 
even if hypothesized hydrogen bonding patterns and definite preferences have been actually found in experimentally 
solved structures \cite{Pabo:1992}.
Moreover regulatory proteins are known to recognize specific DNA sequences directly through atomic contacts between 
protein and DNA and/or indirectly through water-mediated contacts and conformational changes 
\cite{Harrison:1990,Pabo:1992,Gromiha:2004}.
The degree of redundancy and flexibility seems to suggest that the recognition mechanism is ambiguous, therefore the 
prediction of DNA target sequences is not straightforward \cite{Kono:1999}.

DNA protein interactions can be studied using several different computational
methods, which could offer several advantages compared to the current
experimental methods, more laborious and slow. In the following we will indicate, for simplicity, DNA-binding protein 
target sequences with the more specific term ``transcription factor binding sequences'', although the first term is more 
general.\\
Computational tools for the identification of Transcription Factors (TF) binding sequences can be organized in two 
main approaches:
\begin{itemize}
\item
``sequence based methods'' in which a central role is played by the statistical properties of the base
distribution in the DNA regions
which are expected to be involved in transcriptional regulation (see \cite{i1,i2} for a general review on the subject).
\item
``structure based tools'' which use the structural information on protein-DNA complexes derived from X-ray crystallography
 and Nuclear Magnetic Resonance.
\end{itemize}

The main focus of this paper is on the second approach, although the best results will likely be obtained by tools able 
to combine in a clever way these two approaches.

\vskip 0.3cm
\noindent
{\bf Sequence based methods}\\
This type of algorithms can in turn be divided into two broad groups:
\begin{description}
\item{i)}
 enumerative methods,
which explore all
possible motifs up to a certain length (see e.g.
\cite{i10,i11,i12,i13,i14,i15,i16,i17}). 
\item{ii)} local search
algorithms, including expectation maximization and various
flavours of Gibbs sampling (see e.g. \cite{i18,i19,i20,i21}).
\end{description}
It is important to stress that this type of studies cannot be based exclusively on the statistical
features of the DNA regions presumably involved in transcriptional regulation, but must be complemented
with independent information about gene regulation. In this respect three important sources of information may
be used: the functional annotations collected in public databases, gene expression data on a global scale, and the
so called 'phylogenetic footprinting'. In particular this last approach, thanks to the increasing number of
sequenced genomes, has proved to be very effective in these last few years (see e.g.
\cite{i3,i4,i5,i6,i7,i8,i9,i9a,i9b,i9c}).

The major problem of all these tools is the large number of false positives, above all in the case of higher
 eukaryotes (for a thorough analysis of this problem see the interesting assessment of TF binding sites discovery tools 
reported in~\cite{tompa05}). It is exactly to cope with this type of problem that it could be important to resort to 
structure based approaches.

\vskip 0.3cm
\noindent
{\bf Structure based methods\\}

These methods can be broadly divided into two classes according to a nomenclature 
adopted in the context of protein structure prediction \cite{Lazaridis:2000}:\\ 
i) those based on  knowledge based potentials (mostly statistical effective energy functions, SEEFs);  \\
 ii) those based on physical potentials (or physical effective energy functions, PEEFs). \\
SEEFs are energy functions derived from a dataset of known protein-DNA structures.
A set of features is selected (e.g. nucleotide-amino acid contacts, roll angles for DNA bases, interatomic distances, etc.); the process often involves parameter choices, like threshold on distances or interval binning. The statistical properties of these features are compared with a-priori expectations and log-odd scores are derived \cite{Sippl:1990,Lustig:1995,Kaplan:2005,Mandel-gutfreund:1998,Liu:2005,Zhang:2005,Kono:1999}.
At the most basic level, structures may be used to define contacts among DNA bases and protein amino acids and, for each pair of positions, the occurrences of nucleotides and amino acids contacts are used to derive effective potentials \cite{Mandel-gutfreund:1998}. Moreover a statistical potential, taking into account contact geometry and spatial arrangement of contacting residues can be derived \cite{Kono:1999}. Recently interesting developments of this approach have been proposed (\cite{Zhang:2005}\cite{Kaplan:2005}\cite{Gromiha:2004}\cite{Olson:1998}).
The approach suffers from theoretical and practical problems. From the theoretical point of view potentials of mean force are not in general additive and the exact modelization of a-priori expectations (or so-called reference state) may be difficult for complex systems (see e.g. \cite{Zhou:2002}).
The main practical problem is the requirement of a large number of sequences or binding experimental data since the available data may be biased towards specific classes of protein-DNA complexes. 
Moreover datasets generally do not contain unfavourable interactions between amino 
acids and bases since they entail protein-DNA complexes that occur 
naturally. Thus the statistical potential may predict correctly the wild type 
targets as opposed to incorrect ones, but it may not be as good at 
distinguishing among mutants.\\ 
Notwithstanding all caveats usage of SEEFs are widespread in the field of structural predictions. Provided that sufficient data are available these methods are reasonably fast and accurate, as demonstrated for instance in the field of protein structure prediction (see e.g \cite{Baker:2005}). \\
A more radical approach is to estimate the free energy of binding 
starting directly from the available (or homology built) protein-DNA complexes using physical effective 
energy functions (PEEFs). 
This approach has been successfully used in many contexts, ranging from
estimation of DNA- or protein-ligand binding free energy to estimation 
of protein-DNA binding free energy (see e.g. \cite{Wang:2001,Kollman:2000}).
There are, however, many problems connected with the approach which are mainly due to:\\ 
i) difficulties in estimating entropic effects;\\ 
ii) difficulties in properly estimating solvation effects;\\ 
No consensus has emerged on the choice of parameters (e.g. inner dielectric
constant, surface tension coefficient, forcefield parameters) and on the protocols that should be applied;\\
iii) difficulties in estimating gas-phase energy with available forcefields 
which are derived from the analysis of small compounds at equilibrium and do 
not take into account electrostatic polarization.\\
In order to get rid as far as possible of all these problems, binding free 
energies are expressed relative to a reference system and in most computational
 studies optimal parameters have been chosen for matching experimental data.\\
As far as protein-DNA complexes are concerned attempts to compute binding free energies using physics based approaches have started in the 1990s. The electrostatic component of the binding free energy has been studied according to continuum methods and its dependence on temperature and salt concentration has been computed \cite{Zacharias:1992,Misra:1994,Fogolari:1997}. Integration of electrostatics with other components including DNA conformational free energy has been extended from DNA-ligand complexes \cite{Baginski:1997} and protein-peptide complexes \cite{Froloff:1997} to protein-DNA complexes \cite{Misra:1998}. 
Recently Wojciechowski et al. \cite{Wojciechowski:2005} studied the complex of telomerase end binding protein with single stranded DNA optimizing the weights of different contributions in order to reproduce binding data. 
 The availability of the successful analytical generalized Born model treatment of electrostatics solvation effects enabled computation of binding energies with hybrid molecular mechanics/Generalized Born surface accessibility methods by Jayaram et al. \cite{Jayaram:1999}. 
The group of Kollman developed the molecular mechanics/Poisson Boltzmann surface accessibility (MM/PBSA) methodology and applied it extensively to biomolecular systems (see for a review of these applications \cite{Kollman:2000,Wang:2001} and \cite{Kollman:2000,Gorfe:2003,Kollman:2000,Wang:2001} for important extensions of these ideas).\\
However, when MM/GBSA or MM/PBSA energy versus time plots are presented for explicit solvent molecular dynamics simulation snapshots, fluctuations in the range of tens to hundreds of kcal/mol are found, thus posing an issue on the reliability of averages. 
In this respect SEEFs appear much more robust energy estimation methods.\\
In a few very recent reports interesting results have been reported concerning the capability of hybrid methods to predict protein-DNA binding sites \cite{Endres:2004,Oobatake:2003,Oobatake:1993,Havranek:2004,Morozov:2005}. 
In this paper we focus on the application of PEEFs to a single DNA binding protein in complex with many different DNA sequences.\\
The availability of high resolution X-ray 
crystal structure \cite{Beamer:1992} and suitable experimental data makes the $\lambda$ repressor-operator 
complex an interesting system for computational analysis of protein-DNA 
interaction.\\ 
The bacteriophage $\lambda$ repressor protein is a small, 92 amino acid, 
protein that binds the DNA as a dimer. Each monomer binds to an operator half 
site. The amino-terminal domain of $\lambda$ repressor is responsible for DNA binding 
and the carboxy-terminal domain is primarily responsible for dimerization \cite{Johnson:1981}. 
Each monomer contains a typical  helix-turn-helix motif found in a variety of DNA binding proteins 
\cite{Harrison:1990,Brennan:1989}. The free 
energy of binding of $\lambda$ repressor for wild-type $O_R1$ operator DNA and of all 
possible single base-pair substitutions within the operator have been 
experimentally measured using the filter binding assay technique and changes in 
the free energy of binding caused by the mutations have been determined \cite{Sarai:1989}.\\
Besides being a perfect playground to test our methods, the so called ``$\lambda$-switch'' in which the
$\lambda$ repressor is involved is very interesting in itself (for a review see ~\cite{Ptashne:1992}).
This "genetic" switch is tightly regulated by the $\lambda$ repressor and the $Cro$
proteins. In these last years this system, due to its relative
simplicity and to the availability of rather precise experimental data attracted a lot of interest and 
various models (see for instance \cite{Ackers,Shea,Aurell} and references therein)
 have been proposed to describe its behaviour.
Despite these efforts in all these models there are still a few open problems which need to be understood. 
In particular it has been recently realized that in order to ensure the remarkable stability of the $\lambda$
switch one should require a very high non-specific affinity both for the $\lambda$ repressor and for 
$Cro$~\cite{Aurell,Bakk}. Such a prediction is very
difficult to test experimentally but could rather directly be evaluated with the tools which we shall discuss in
this paper. In fact one of the main goal of the test which we shall perform on the $\lambda$ repressor will be
the evaluation of its non-specific binding energy and the comparison with the prediction of the model discussed
in~\cite{Bakk}.

In the present work we apply
different techniques to evaluate the binding affinities by means of 
computational methods. 
It is assumed that the relative free energy of binding of a protein to different DNA sequences may be expressed as the sum of a molecular mechanical term, that 
includes the non-bonded electrostatic and Van der Waals contributions, and 
a hydration term that can be further split in a 
polar and a hydrophobic contribution. 
Due to the peculiar nature of hydrogen bonds similar alternative models are tested where an energy term proportional to the number of hydrogen bonds is added.\\
The systems studied here differ only in one or two base-pairs and therefore the inaccuracies implicit in the assumption of rigid docking, of the solvation model, of the treatment of entropy and in lack of a complete conformational search for side chains at protein-DNA interface should mostly cancel out in comparison.\\
The aims of this paper are:\\
1) to provide an assessment of the accuracy of different methods and protocols by comparison with experimental data;\\
2) to provide a reliable estimate of non-specific binding energies;\\
3) to propose a protocol for the prediction of DNA-binding target sequences which makes no use of sequence information.\\
To pursue these objectives we make use of extensive computations and 
address several specific issues. 
In particular:\\
i) we estimate optimal weights for different contributions to
DNA-protein binding free energies using different solvation models; \\
ii) for 52 single base-pair mutants we perform 1 ns molecular dynamics (MD) runs and we assess the effect of MD on the computed binding energies;\\
iii) we compute MM/GBSA binding energies for one thousand complexes where 
the bases of the double stranded DNA are substituted according to randomly 
generated DNA sequences in order to estimate non-specific binding free energy;\\
iv) we scan the entire bacteriophage $\lambda$ genome with the scoring profiles obtained from free energy computations. One of the profiles is obtained making use only of the structural data available for a single molecular complex, with no sequence information.\\
The statistical analysis of the results show that computational methods may offer a predictive tool truly complementary to sequence-based identification of DNA-binding protein target sequences. This is particularly important in view of the emergence of consensus protocols where the independence of the different methods is a prerequisite.\\

\section*{Results and discussion}
 
Binding free energy changes between the $\lambda$ repressor dimer and the DNA operator mutants have been calculated
 using different methodologies, as described in the {\sl Methods} section. 

\subsection*{MM/DDDC-OONS}

We calculated the binding free energy between the $\lambda$ repressor dimer and the DNA operators, after having 
energy-minimized every complex using a distance dependent dielectric constant (1r, 2r, 4r, 8r, respectively, in order to match subsequent energy evaluation).
The interaction energy between the protein and the DNA, $\Delta U(\vec r_1,\ldots,\vec r_n)$, has been evaluated
using four values for $\epsilon$ (1r, 2r, 4r, 8r) then the solvation term $\Delta G_{solv}$ has been 
determined according to the model of Oobatake et al. \cite{Oobatake:2003} using Eq. \ref{eq2}.
The best scaling factors have been determined (together with the standard deviation computed according to Eq. \ref{sigma2}) fitting the set of experimentally measured protein-DNA binding affinities and are reported in Table 1.
The addition of a specific hydrogen bond term reduces the coefficients of the electrostatic term. 
The RMSD and the correlation coefficient $r$ have been computed and a 
leave-one-out scheme has been adopted, in order to verify the performance of the model (Table 2).
The same analysis has been performed for 5000 replicas of the dataset with one third of the set left out and used for cross-validation. The average RMSD and correlation are essentially the same reported for the leave-one-out scheme reported in Table 2.
From the same analysis variances of the coefficients have been estimated with essentially the same results as those reported in Table 1.\\
The best correlation coefficient ($r$ = 0.703 for MM/DDDC-OONS(+HB) model) has been obtained for $\epsilon=4r$, although values of $\epsilon=2r$
 and $\epsilon=8r$ gave very similar results for both MM/DDDC-OONS and MM/DDDC-OONS(+HB) models. 
Except for the MM/DDDC-OONS(+HB) model with  $\epsilon=1r$, the F-statistic shows that the model is significant (p $<$ 0.001).
The dielectric constant 
$\epsilon=1r$, which gives the worst results tends in many cases to overestimate binding free energy 
changes lower than 1.0 kcal/mol whereas binding free energy changes greater than 2 kcal/mol are underestimated. 
A similar behaviour has been observed for $\epsilon= 2r, 4r, 8r$, even if these models are able to better 
reproduce binding free energy changes, in particular improvements have been obtained for values lower than 1.0 kcal/mol (Figure 1).\\ 

\begin{figure}[ht]
\begin{center} 
\vspace*{0.1cm}
\includegraphics[scale=0.7, angle=0]{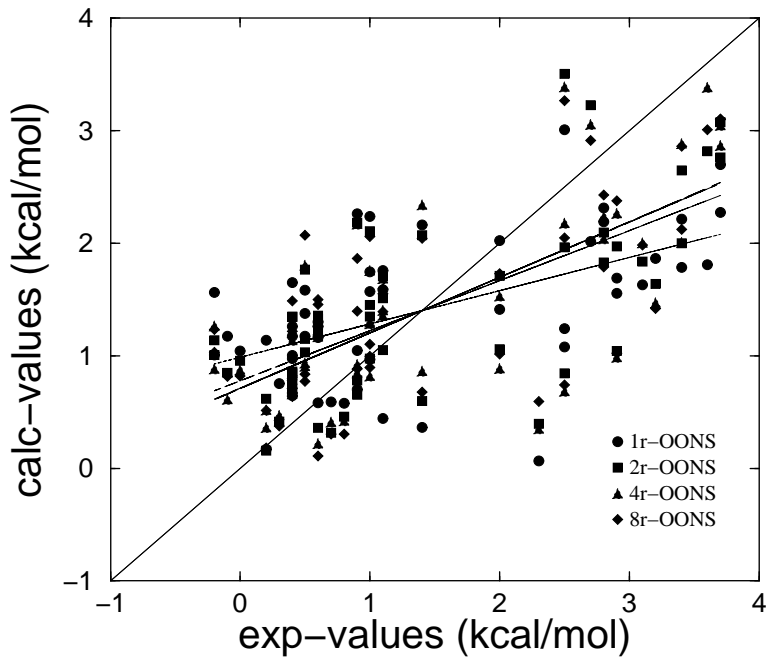}
\vspace*{0.2cm}
\caption{ Computed binding free energies (MM/DDDC-OONS(+HB) model) versus experimental measurements,  using a distance dependent dielectric constant ($\epsilon= 1r, 2r, 4r, 8r$). The correlation coefficients between calculated and experimental values are 0.543, 0.667, 0.703 and 0.701 for $\epsilon= 1r, 2r, 4r, 8r$, respectively.}
\vspace*{-0.5cm}
\end{center}
\end{figure}

The analysis of the best scaling coefficients is not straightforward because there is a strong correlation between the energy terms. 
For instance, for all $\epsilon$ models the electrostatic term is strongly anticorrelated with the OONS solvation term.\\
Moreover the estimated variance of coefficients is often very large. 
Notwithstanding these difficulties it is worth noting that some terms appear to be particularly important. 
For instance each protein-DNA hydrogen bond (when explicitly included in the model) appears to contribute -0.15 to -0.27 kcal/mol, depending on the electrostatic model assumed.\\ As expected the electrostatic term is reduced when hydrogen bonds are taken into account separately. For $\epsilon = 2r $ the best scaling coefficient $x_{DDDC}$ changes from 0.154 to 0.182 upon removal of the term proportional to the number of hydrogen bonds.\\
The correlation between the different contributions is reflected in the changes, with changing dielectric model, of the OONS term scaling factor, which is always strongly reduced by the scaling factor ranging from 0.075 to -0.066.
Finally the constant term which takes into account common entropic terms (which can be estimated to be in the range 20 to 40 kcal/mol) and the free energy of binding of the reference complex (which implies the addition of 11.3 kcal/mol), expected to be in the range of 30 to 50 kcal/mol, is slightly larger than expected. \\

\subsection*{MM/DDDC-HP}

The OONS solvation term is accounting for both apolar and electrostatic solvation terms which should be already taken into account, at least partly, in the distance dependent dielectric constant. The same calculations described above have been performed using a similar approach in which the solvation term of
the binding free energy is taken to be proportional to the polar/apolar accessible
surface area of the molecule (see Eq. \ref{eq6}). The best scaling factors have been determined fitting the 
set of experimentally measured protein-DNA binding affinities (Table 1).
The quality of the computed binding free energies $\Delta G_{calc}$ has been assessed evaluating the linear correlation coefficient $r$ and the 
root mean square deviation (RMSD) between calculated and experimental values. 
In order to verify the performance of the model, a leave-one-out scheme has been adopted (Table 2).
The F-statistic shows that the model is significant (p $<$ 0.001).\\
All the values of the distance dependent dielectric constant which have been tested gave a quite high and 
similar linear correlation coefficient. The highest correlation value ($r$ = 0.745) was obtained for $\epsilon=2r$ for the MM/GBSA(+HB) model
and the lowest ones for $\epsilon=1r$ similar to the MM/DDDC-OONS model. 
Generally, binding free energy changes 
lower than 1.0 kcal/mol are overestimated whereas binding free energy changes greater than 2 kcal/mol are 
underestimated in all the cases. More accurate predictions have been obtained for  $\epsilon=2r$, in 
particular for values lower than 1.0 kcal/mol (Figure 2).\\    
\begin{figure}[ht]
\begin{center} 
\vspace*{0.1cm}
\includegraphics[scale=0.7, angle=0]{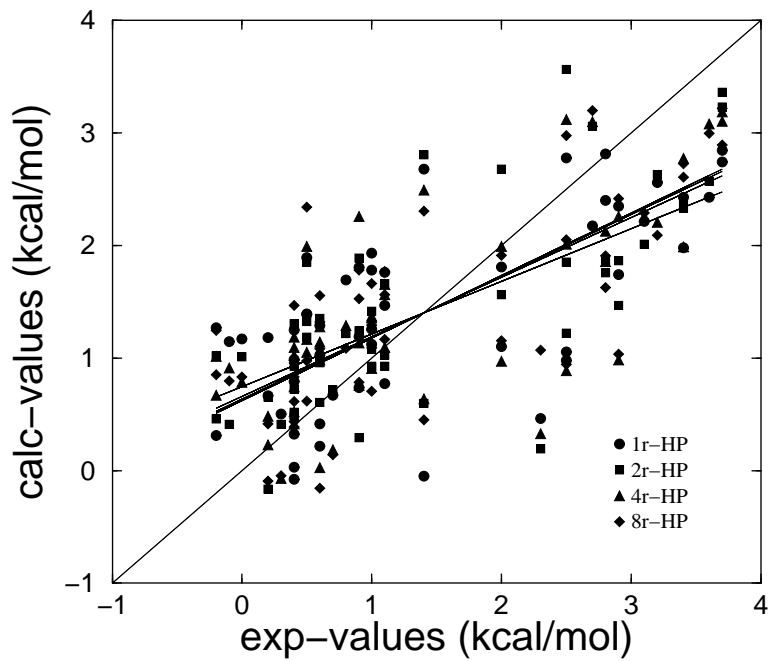}\quad \quad
\vspace*{0.2cm}
\caption{
Computed binding free energies (MM/DDDC-HP(+HB) model) versus experimental measurements, using a distance dependent dielectric constant ($\epsilon= 1r, 2r, 4r, 8r$). The correlation coefficients between calculated and experimental values are 0.684, 0.745, 0.728 and 0.739 for $\epsilon= 1r, 2r, 4r, 8r$, respectively.}
\vspace*{-0.5cm}
\end{center}
\end{figure}
The optimal scaling coefficients are in the expected range (Table 1),
in particular for $\epsilon= 2r, 4r, 8r$ 
the constant term $x_{const}$ is in the range 10-20 kcal/mol, moreover the coefficients $x_H$ and $x_P$ have the right order of magnitude of typically used surface tension coefficients for water biomolecular interface, even if the sign is incorrect. 
It should be noted however that there is a strong correlation (ranging in this case from 0.2 to 0.6) between the coefficients of most terms and the coefficient of the constant term. \\
Also for the present model the addition of an explicit hydrogen bond term
reduces the coefficient of the electrostatic term as could be expected. \\
These results support the conclusion that, in general, there is no advantage in using the detailed solvation models compared to the simpler polar/apolar model, as far as the binding free energy is concerned. 
Based on the range of the scaling coefficients the two models appear of similar quality.
Scaled free energy components for the MM/DDDC-HP(+HB) model are reported in Table 3.

\subsection*{MM/GBSA}

In this approach all structures have been energy-minimized using the Generalized Born solvent model, then the binding 
free energy for every molecule has been calculated according to the MM/GBSA model using the Eq. \ref{eq7}.
As in the previous cases, we determined the best scaling factors (and standard deviations according to Eq. \ref{sigma2}) fitting the set of experimentally measured protein-DNA binding affinities (Table 4),
then we assessed the quality of $\Delta G_{calc}$ predictions evaluating the linear correlation
 coefficient $r$ and the root mean square deviation (RMSD) between calculated and experimental values. 
Finally we 
verified the performance of the model, using the leave-one-out scheme (Table 2).
The same analysis has been performed for 5000 replicas of the dataset with one third of the set left out and used for cross-validation. The average RMSD and correlation are essentially the same reported for the leave-one-out scheme reported in Table 2. The standard deviations of the coefficients are essentially the same as reported in Table 4.
Our calculation shows that the MM/GBSA(+HB) model gives the best performances ($r$ = 0.746), although the linear correlation
coefficient between calculated and experimental values differs slightly from the best values obtained from the other models. 
The F-statistic shows that the model is significant (p $<$ 0.001).\\
Computed values versus experimental data are reported in Figure 3. 
\begin{figure}[ht]
\begin{center}
\vspace*{0.1cm}
\includegraphics[scale=0.7, angle=0]{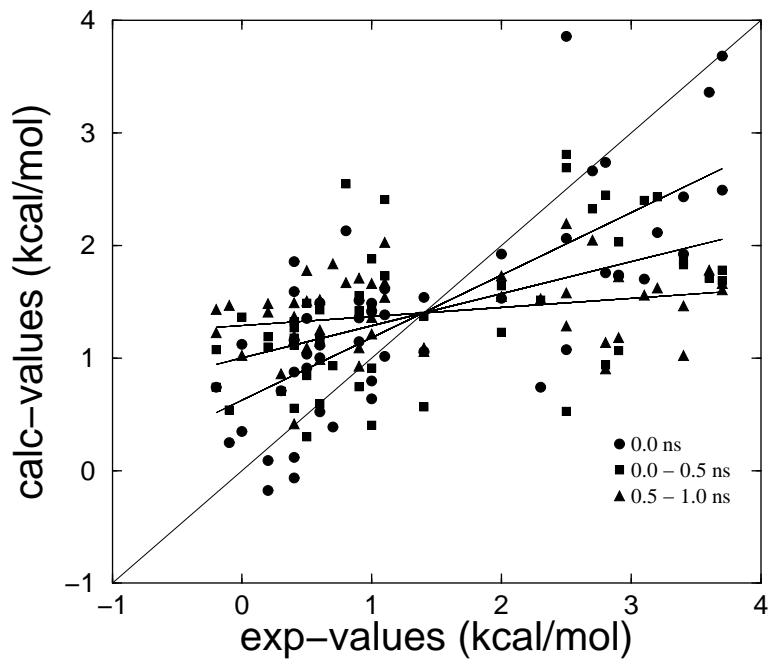}\quad \quad
\vspace*{0.2cm}
\caption{ Binding free energy values calculated using the MM/GBSA(+HB) model versus experimental values 
Structures at 0.0 ns refer to the minimized complexes. The other two sets of data have been obtained by averaging over the MD simulation times 0.0 to 0.5 ns and 0.5 to 1.0 ns.
The correlation coefficients between calculated and experimental values are 0.746, 0.534 and 0.284 for the minimized complexes, for the averages over time 0.0 to 0.5 ns and for the averages over time 0.5 to 1 ns, respectively.}
\vspace*{-0.5cm}
\end{center}
\end{figure}   
As far as the scaling coefficients are concerned (see Table 4),
it is worth noting that addition of an explicit hydrogen bond term has a dramatic effect on the coefficients of van der Waals and electrostatic terms, as could be expected, because the latter terms already take into account hydrogen bond energetics. 
For the MM/GBSA model (with no explicit term for hydrogen bonds) the 
coefficients of the electrostatic and GB solvation terms are 0.16 and 0.14 
which correspond to a dielectric constant of $\sim 6$. 
Surface tension coefficients $x_P$ and $x_H$  (-0.010 and -0.029 respectively) have the same order of
magnitude of the commonly used surface tension coefficient (ca 0.02 kcal/mol \AA$^{-2}$), but opposite sign. However the terms proportional to the solvent accessible surface area are strongly correlated to each other and to the constant term.\\
The constant term is -11.7 kcal/mol, lower than what expected, probably as a consequence of the correlation of this term with the polar and hydrophobic surface area terms (the linear correlations of the coefficients are 0.51 and 0.75, respectively). The standard deviation of this term is however very large (28.0 kcal/mol).\\
Scaled free energy components for the MM/GBSA(+HB) model are reported in Table 5.

The analysis of Table 5 shows that the most important feature for computing the binding
free energy is the number of intermolecular hydrogen bonds. 
The correlation of the associated energy term with the experimental free energy of binding 
is 0.58. Other terms are strongly correlated among each other and therefore it is difficult to
single out specific contributions. 
The correlation between different energetic terms range 
from -0.99, for GB solvation energy and Coulombic energy, to 0.44, for
GB solvation energy and polar area burial energy term.\\
18 single base-pair mutants exhibit large (greater than 2.0 kcal/mol)
unfavourable free energy of binding.
Loss of hydrogen bonds contributes for 1 or 1.5 kcal/mol for mutants
14CG, 8AT, 8GC, 9AT, 18AT, 12TA, 12GC, while for other mutants the most
important unfavourable contributions come mostly from Coulombic and van der Waals
terms. 
It should be noted, however, that solvation terms are correlated with Coulombic and van der Waals terms. \\
This analysis is in general in line with the detailed analysis reported by Oobatake et al.
\cite{Oobatake:2003}, although the exact values of energy contributions differ. 

\subsection*{Analysis of molecular dynamics trajectories}

The procedure used for computing binding energies may suffer from incomplete relaxation and incomplete conformational sampling. An approach that has been used in the past for sampling more conformations and reduce the effect of fluctuations is to analyse snapshots from molecular dynamics runs. In many studies no scaling factor was applied at all, with good results.\\
We performed 1 ns of MD simulations for every structure in order to test the effectiveness of a first principles computation of binding free energies and to check the effect of molecular dynamics relaxation on the computed energies. 
We calculated the average value of every component of the binding free energy using snapshots taken every 50 ps, then we used the same set of fitting equations using average values to determine the best scaling factors.  
We chose to use the MM/GBSA(+HB) model for computing binding free energies because it gave good results on the starting structures and the coefficients can be used to monitor the quality of the fitting.
Figure 3 
reports the MM/GBSA(+HB) binding free energy values obtained from MD simulations versus experimental values.
The quality of the computed binding free energies $\Delta G_{calc}$ has been assessed evaluating the linear correlation coefficient $r$ and the root mean square deviation (RMSD) between calculated and experimental values 
(see Table 6).
Results show that MD simulations do not improve the prediction capabilities of the model. Actually the linear correlation coefficient
 calculated averaging over 1.0 ns is 0.356, much lower than the correlation at t = 0.0 ns. Results obtained averaging over the time interval 0.0-0.5 ns, gave a linear correlation coefficient comparable to what obtained with other models on the starting structures (r = 0.534) but lower than the linear correlation coefficient obtained at t = 0.0 ns. The linear correlation coefficient between experimental values and the results obtained averaging over the time interval 0.5-1.0 ns, is $r$ = 0.284, indicating that MD causes the loss of any correlation.\\
Moreover optimal scaling factors obtained averaging over the time interval
0.5-1.0 ns have the tendency to lose any physical meaning (Table 7).
When optimal scaling factors obtained on the starting structure are used to compute binding free energies using average values, no correlation is detectable with experimental data.
The value of the binding free energy change of every complex across 1 ns of simulation has been observed to strongly fluctuate, making it difficult to obtain an accurate estimate of it.\\
In order to verify whether this problem could have been circumvented using a larger conformational sampling, the simulations of 10 mutants have been extended to 4 ns, obtaining a total of 400 snapshots for every simulation. In particular we extended the simulations of the wild type complex and the best and the worst mutants (G17-C25 and T14-A28 respectively) with negative results. 
Although the system is most probably not fully equilibrated, it is reasonable to suspect that even longer (in the range of few tens ns) molecular dynamics simulations will not improve the results obtainable on the starting structures. \\
The main reasons of failure of this approach are probably the large conformational fluctuations developing in MD simulations and  the combination of relatively short molecular dynamics simulations with snapshots energy evaluation using the MM/GBSA(+HB) continuum model.
Large conformational fluctuations observed in MD simulations are reflected in energetic fluctuations in the range of tens of kcal/mol, thus posing an issue on the reliability of the free energy average values. Moreover, since we observed that the results could not be improved extending the simulation time, it is reasonable to ascribe the failure of the method, at least partially, to inaccuracies in the force field parametrization. Actually, all force fields are based on numerous approximations, in particular nucleic acid force fields could suffer from two main problems which could give rise to inaccuracies. The first is that the target experimental data used in the optimization process are typically crystal structures of DNA and RNA. However, the presence of the lattice environment in crystals is known to influence the structure of DNA, limiting the transferability of crystal data to solution. The second
is the treatment of electrostatics which is crucial in these simulations, given the polyanionic nature of DNA. In particular,  the electrostatic polarization, which is an effect that can significantly reduce electrostatic interactions of partial atomic charges, is very important for accurate treatment of interactions in different environments, since significant structural changes of DNA may occur in response to environment.

\subsection*{Correct predictions}

Table 8 shows the number of correct predictions, according to the criteria described in the {\sl Methods} section. \\
In the last column the number of cases in which the 
difference $D=|\Delta G_{exp}-\Delta G_{calc}|$ is lower than 0.3 kcal/mol, that is the number of the more accurate predictions, has been reported.
It should be noted that the fitting of coefficients aims at minimizing the RMSD between calculated and experimental values and not at maximizing the number of ``correct'' predictions. When a simple simulated annealing procedure is applied to the coefficients the number of correct predictions can be increased by several units.
It is instructive for instance to consider the MM/GBSA(+HB) model, where 41 ``correct'' predictions can be achieved with minor (mostly less than 10\%) variations relative to the starting values of coefficients. \\

From this qualitative point of view, the prediction capabilities of the different models can be compared. The best performing models appear to be the MM/DDDC-HP model with $\epsilon = 2r$. On average the DDDC-HP model performs better than the similar DDDC-OONS model. For $\epsilon = 1r$ results are worst than for higher $\epsilon$ values.\\
Molecular dynamics trajectories were analysed similarly, using average values for the different contributions to the free energy of binding.
In particular the lowest number of correct 
predictions has been obtained averaging over the time interval 0.5-1.0 ns, actually there is no cases in which both $\Delta G_{exp}$
 and $\Delta G_{calc}$ are $<$1.0 kcal/mol and the number of cases in which $\Delta G_{exp}$ and $\Delta G_{calc}$ are 
separated by less than 0.3 kcal/mol has been strongly reduced. \\
Generally, we observed that the number of cases in which
 $\Delta G_{exp}$ and $\Delta G_{calc}$ are both $<$1.0 kcal/mol decreases while the number of cases in which $\Delta G_{exp}$ 
and $\Delta G_{calc}$ are both $>$ 1.0 kcal/mol remains nearly constant; however at the same time the number of cases in which $\Delta G_{exp}$ 
and $\Delta G_{calc}$ are separated by less than 0.3 kcal/mol strongly decreases, indicating that there is a reduction of the accuracy in reproducing experimental binding energies.
Overall this analysis is consistent with the analyses reported in the previous sections.

\subsection*{Validation of the MM/GBSA model using the Cro-OR1 complex}

The optimal scaling coefficients are likely to depend on the complex
and mutants studied. In order to verify that such coefficients do not  
produce wild results when applied on different complexes with similar binding 
features we considered the $Cro$ OR1 complexes which was obtained
from crystallographic structure (PDB id. code: 6CRO) after mutation of
14 bases. The $Cro$ protein belongs to the same family of $\lambda$ repressor but to a different domain, according to SCOP classification \cite{Scop} and it has very limited similarity with $\lambda$ repressor
although they bind DNA in a similar fashion. This system is therefore suited 
for testing the overall quality of the scaling procedure.
Also for $Cro$ a set of measurements for each mutant of the OR1 sequence is available \cite{Takeda:1989}. 
When all contributions to the binding free energy, computed according to the MM/GBSA(+HB) model, are scaled by the coefficients determined on the $\lambda$ repressor complexes the computed energies show a remarkable correlation coefficient of 0.62 with the experimental values, although the binding energies are overestimated by approximately 10 kcal/mol. This fact could reflect differences in the entropic contribution to binding (arising from restriction in side chain and backbone mobility) that are likely to be different for the two systems. Indeed, the crystallized $Cro$ protein is roughly only two thirds of the repressor sequence.
Notwithstanding the differences in overall binding energy, the binding differences for the mutants are on average reproduced by the energetic model. \\
As a further test, we performed the reverse analysis where the scaling coefficients are obtained on the $Cro$-OR1 complex and validated on the $\lambda$ repressor-operator complex. Also in this case the computed energies show a remarkable correlation coefficient of 0.69 with the experimental values, although they are all underestimated by approximately 16 kcal/mol.\\

In order to verify how sensitive the scaling coefficients are to the experimental data used in the fit, we calculated the binding free energies of 
$Cro$ and each mutant of the OR1 sequence according to the MM/GBSA model, using Eq. \ref{eq7}. Finally we combined the two experimental datasets 
of $\lambda$ repressor and $Cro$ and we refitted the model.\\
As in the previous cases, we calculated the best scaling factors fitting the set of experimentally measured protein-DNA binding affinities (Table 4),
then we assessed the quality of $\Delta G_{calc}$ predictions evaluating the linear correlation coefficient $r$ and the root mean square deviation 
between calculated and experimental values. Finally we verified the performance of the model, using the leave-one-out scheme.
The best performance has been obtained for the MM/GBSA(+HB) model, which gives a correlation coefficient $r$ of 0.69 and a rmsd of 0.74 for $Cro$ and 
a correlation coefficient $r$ of 0.67 and a rmsd of 0.83 for the two combined systems.  
The same analysis has been performed for 5000 replicas of the dataset with one third of the set left out and used for cross-validation. 
The average RMSD and correlation are essentially the same reported for the leave-one-out scheme.
From the same analysis variances of the coefficients have been estimated with essentially the same results as those reported in Table 4.
As far as the scaling coefficients are concerned (see Table 4), by comparing the results obtained for $\lambda$, $Cro$ and the two combined systems, we can observe that the sets of values obtained for $\lambda$ and $\lambda + Cro $ are all in the same range except for the constant term, probably as a consequence of the fact that the entropic contribution to binding are likely different for the two systems. However it is worth noting that the standard deviation of this term is very large in both cases. As far as the scaling coefficients obtained for $Cro$, they are rather different from the others, except for $x_{vdw}$ and $x_{HB}$, which scale the Van der Waals and H-bonds contributions respectively. However we observed that the electrostatic and GB solvation terms are strongly correlated to each other (the linear correlation coefficient is 0.998), as well as the constant term and the polar and hydrophobic surface area terms (the linear correlation of the coefficients is 0.645 and 0.784). The standard deviation of the constant term is also very large (see Table 4). 
Overall these results validate the approach for predicting binding free energies for similar protein-DNA complexes.\\

\subsection*{Analysis of non-specific protein-DNA binding}

In order to study non-specific protein-DNA binding one thousand random DNA sequences have been generated and each sequence has been threaded onto the DNA phosphate backbone of the crystal structure in order to obtain a set of structural models with new DNA sequences. 
Minimization was performed according to the protocol described in the {\sl Methods} section. We refer to to this set of complexes as to the ``non-specific'' set.\\
Binding free energies for each member of the generated non-specific set have been computed according to the MM/GBSA(+HB) model, using the optimal scaling factors determined by fitting the 52 experimental data (see Table 4).\\
We calculated the Z-scores of both the random structures and the single base-pair mutants, i.e. the distribution of the difference between the binding free 
energy of a complex and the average energy of the non-specific set, normalized by the standard deviation of the computed energies. Z-scores 
represent the specificity of a complex, with larger negative values corresponding to higher specificity.   
Figure 4 
shows the distribution of computed energies. The distribution of the Z-scores of the single base-pair mutant complexes, is found at the negative tail of the non-specific distribution, indicating that these complexes are more stable than the complexes 
formed with a DNA random sequence, as one expects. 
\begin{figure}[ht]
\begin{center}
\vspace*{0.1cm}
\includegraphics[scale=0.7]{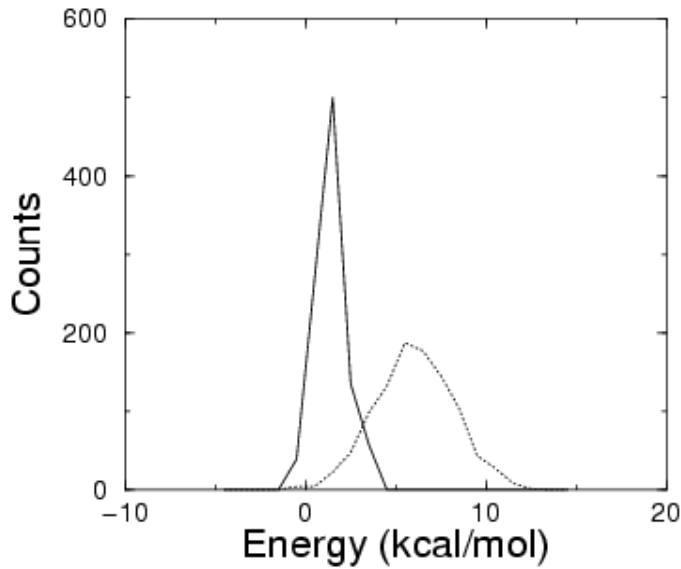}\quad \quad
\vspace*{0.2cm}
\caption{Distributions of the calculated binding free energy (bin width = 1 kcal/mol) for the ``non-specific'' set (dotted line) and for the single base-pair mutants (continuous line). The distributions are normalized to the same total number of counts.}
\vspace*{-0.5cm}
\end{center}
\end{figure}
The computed energies have an average difference of 4.8 kcal/mol and a standard deviation of 2.2 kcal/mol, giving thus an average z-score for the single base-pair mutants of 2.14 and 2.87 for the lowest computed energy in the set.\\
The average non-specific binding energy seems surprisingly low (meaning that it implies that a rather large fraction of $\lambda$ repressors present 
in the cell is actually non-specifically bound to DNA) but, remarkably enough, it agrees within the errors with the value 
proposed in~\cite{Bakk} as a way to explain the impressive stability of the $\lambda$-switch.\\
It is interesting to compare the computed free energies of binding for the non-specific DNA complexes with those expected based on single mutants binding energies under the assumption of additivity. 
The expected free energies are higher than those computed by optimal scaling of contributions. The average difference, with respect to the specifically bound sequence, are 18.4 kcal/mol and 6.1 kcal/mol, respectively. 
This has been interpreted as a consequence of the fact that adjacent multiple substitutions may introduce additional energy minima compared to single mutations in a tight complex. This result is in line with the saturation effect in observed vs. predicted binding energy that has been described by Stormo and co-workers \cite{Benos:2000,Benos:2002} and recently experimentally demonstrated \cite{Maerkl:2007}. 
It is also interesting to note that the non-specific binding energy is comparable to the energy computed by Northrup and co-workers for loosely docked complex of $Cro$ to non-cognate DNA \cite{Northrup:1997}, which implies that the mode of binding may substantially change for non-specifically bound DNA sequences.
This would be consistent with the capability of the protein of sliding along DNA, which would not be feasible for a tight complex.

\subsection*{Identification of putative transcription factor binding sites}
 
The aim of this section is to understand whether the methods described here can be used for searching genomes for candidate transcription factor binding sites.\\
In particular we aim at verifying:\\
i) whether the MM/GBSA(+HB) model is able to identify transcription factor binding sites in the absence of thermodynamic data about single base-pair mutants, but just knowing the recognized sequence;\\
ii) whether some predictions can still be afforded in the absence of thermodynamic data and of any information on recognized sequences. The latter situation could be encountered when a model of the complex is built by homology and differences in protein DNA-contacting residues imply a different specificity.\\

\subsubsection*{Identification of putative transcription factor binding sites knowing the bound sequence}

The analysis in the previous section used knowledge about single base-pair mutants which is rarely available.
Here we ask what predictions can be made when no thermodynamic data on mutants or wild-type sequence binding is available, but the cognate sequence is available. 
One thousand random DNA sequences were generated and the corresponding structural models were built by performing mutations on the double stranded DNA in the complex crystal structure using the program WHATIF \cite{Vriend:1990}. 
Structures were energy minimized using the same protocol used for the MM/GBSA(+HB) methodology. Assuming that random sequences will have a larger free energy of binding compared to the bound sequence, optimal scaling parameters were sought in order to make the free energy difference in binding with respect to the naturally occurring complex equal to 10.0 kcal/mol. This value is arbitrary, albeit not unrealistic.
Eq. \ref{eq7} is solved (in a least square sense) subtracting the row corresponding to the wild-type complex from all other rows, and fixing all the energy differences equal 10.0. The differences in Coulombic, van der Waals, GB solvation energy, polar and apolar surface area and number of hydrogen bonds, with respect to wild-type complex, have been tabulated and the optimal scaling parameters have been determined.\\ 
The free energies computed on the random sequences have been used to compute single base-pair mutant free energies as described in the {\sl Methods} section. The single base-pair mutant energies for the wild type sequence have been reset to 0.0 (this assumes that the specific bound sequence is known) and the lowest computed single base-pair mutant binding energy has been subtracted from all other values.\\
The plot of computed single base-pair mutant energies vs. experimental energies (computed under the hypothesis of additivity) shows a good correlation (0.58) but seems insufficient for predictive purposes.
However, when the bacteriophage $\lambda$ genome is scanned using the corresponding free energy matrix (see Methods), high-affinity binding sites are correctly recognized, and in general the energies computed using the matrix and those predicted based on addition of single base-pair mutation effects are well correlated (corr. coeff. 0.74).\\
We asked what is the advantage of such computation compared to the simpler model that assigns a constant energy penalty to each mutation over the specific bound sequence. In such case the correlation between the computed and reference binding energies is slightly lower, but still significant (0.72). 
The advantage of using computational results over a much simpler single parameter approach seems therefore very limited, although the 1\% best sites predicted by the MM/GBSA(+HB) energy and the simple mutation models display only 15\% common sites, proving that the two methods are largely uncorrelated.\\

\subsubsection*{Identification of putative transcription factor binding sites without knowing the bound sequence}

As a last test we simulate a realistic situation in which no thermodynamic data or information on the recognized sequences are available. We considered the set of one thousand random DNA sequences and the corresponding structural models built by performing mutations on the double stranded DNA in the complex crystal structure as the only information available. Obviously the crystallographic complex does contain information on the specific sequence because protein and DNA conformations are fitting each other in the complex. 
If non-specific complexes were to be built by homology without knowing the exact DNA sequence bound, it is likely that side chains would be placed differently with different results. Finally, structures were energy minimized using the same protocol used for the MM/GBSA(+HB) methodology.
As in the tests above we found optimal scaling factors in order to make all (non-specific) binding free energies equal to 10.0 kcal/mol. \\
In order to avoid a trivial solution to the fitting problem with all coefficients equal 0.0 except the constant term equal 10.0, we follow a two-step procedure. 
In the first step we assume a reasonable value (30 kcal/mol) for the constant term which must be brought to the left-hand side of Eq. \ref{eq7}.
Coulombic, van der Waals, GB solvation energy, polar and apolar surface area and number of hydrogen bonds have been evaluated, Eq. \ref{eq7} is then solved (in a least square error sense) and the optimal scaling parameters have been determined. The lowest binding energy sequence according to the scaling parameters is determined. The row corresponding to this complex is subtracted from all other rows thus removing the constant term. \\
In the second step the newly obtained matrix, which does not include the constant term anymore is used to find the best coefficients to make all the energy differences equal 10.0 kcal/mol. 
Therefore all energies are expressed relative to the lowest computed energy at the first step.\\
The free energies computed on the random sequences have been used to compute single base-pair mutant free energies as described in the {\sl Methods} section.  At variance with the test performed above we do not set to 0.0 the energies of specific bound sequence (which is assumed here to be unknown). 
The correlation coefficient between computed and experimental energies (computed under the assumption of additivity) for the bacteriophage $\lambda$ genome is 0.50 (Figure 5).\\ 
\begin{figure}[ht]
\begin{center} 
\vspace*{0.1cm}
\includegraphics[scale=0.7, angle=0]{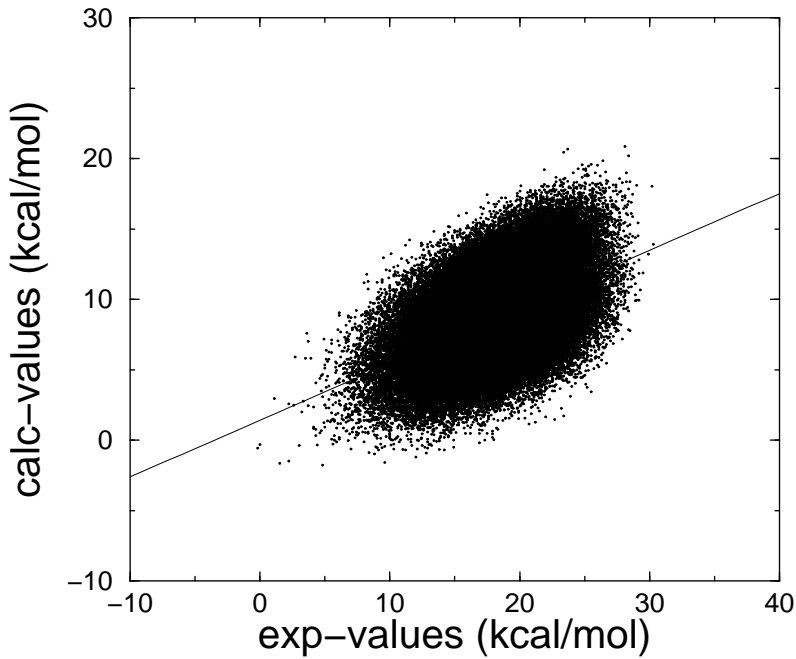}
\vspace*{0.2cm}
\caption{Binding free energies predictions without using specific bound sequence nor thermodynamic information versus binding free energy values obtained under the hypothesis of additivity \cite{Benos:2000,Benos:2002} using experimental data on single base-pair mutants. The correlation coefficient is 0.50.}

\vspace*{-0.5cm}
\end{center}
\end{figure}

As a further test of the performance of the approach we generated the logo \cite{Crooks:2004} of the 10 best binding sequences according to the thermodynamic data on single base-pair mutants and those found with the present approach (Figure 6). An overall agreement between the two logos is apparent.\\ 

\begin{figure}[ht]
\begin{center} 
\vspace*{0.1cm}
\includegraphics[scale=0.7, angle=0]{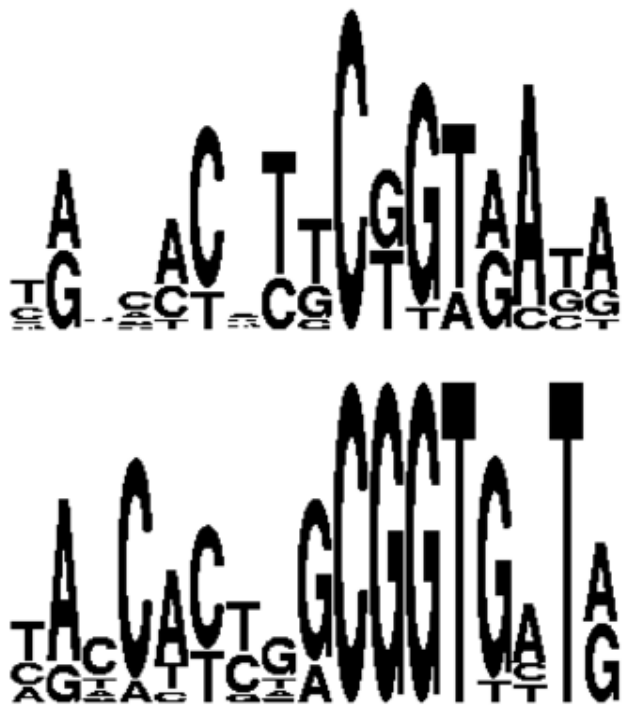}
\vspace*{0.2cm}
\caption{Logos obtained from the ten best binding sequences according to the experimental data of Sarai et al. (ref.\cite{Sarai:1989}) (lower panel) and according to the computations on ``non-specific'' complexes complexes with no sequence or thermodynamic data information (upper panel)}
\vspace*{-0.5cm}
\end{center}
\end{figure}

\section*{Conclusions}

In the present work physical effective energy functions are used to estimate
the free energy of binding of $\lambda$ repressor to the DNA operator and single base-pair mutants, for which thermodynamic data are available.\\
Thermodynamic data allow one to study the best results achievable, with the modeling approach and energy functions presented here, with models that assume that the binding energy is a linear combination of different contributions.\\
Simple models that use a distance dependent dielectric constant and simple terms for surface area proportional energy contributions and for hydrogen bonding perform surprisingly well for values of $\epsilon$ ranging from $2r$ to $8r$.\\
A two-parameter model for surface area proportional energy contributions performs
better than the more complex model of Oobatake et al. \cite{Oobatake:1993}, which was however not derived for usage in the more complex energy functions employed here. \\
The performance of MM/GBSA(+HB) and to a lesser extent MM/GBSA model is comparable to or superior to other models.
A conclusion for the MM/GBSA model is that electrostatic energies should be reduced by a proper scaling factor corresponding to dielectric constants in the range of 6. This conclusion is reached also by a similar analysis of protein $Cro$-operator mutants.\\
The effect of molecular dynamics on the computed binding free energies 
is in general negative and the reproducibility of the experimental values decreases with the increase of simulation time considered. 
This may be a consequence of the large fluctuations developing in MD simulations which probably would require a much longer simulation time. 
Moreover it is reasonable to take into account that the poor performance of the method can be partially caused by the errors in the force field used in MD simulations. 
Another plausible source of inaccuracy is the mismatch between the energy model and system representation used in MD simulation and those used for minimization and energy evaluation.
It appears therefore that it
is worth to invest more time in optimizing the starting structure, rather than for sampling the conformational space by molecular dynamics simulations, or, alternatively, to adopt different strategies for sampling protein and DNA flexibility \cite{Bonvin:2006}.\\
The analysis of non-specific complexes using the best performing energetic model with properly scaled coefficients allows to evaluate a non-specific binding energy difference, with respect to the specific bound sequence, of 6.06 $\pm$ 2.17 kcal/mol, definitely lower than what expected based on an additive model (18.1 kcal/mol for the single base-pair mutants computed energies). This result is in line with the saturation effect described by Stormo and co-workers \cite{Benos:2000,Benos:2002} and with the theoretical analysis of Bakk and Melzer \cite{Bakk}.\\
Although the results presented on single base-pair mutants are not exciting, using computational methods may be very useful for identifying transcription factor binding sites.\\
When no thermodynamic data are available but the specific bound sequence is known the computed MM/GBSA(+HB) free energies are slightly more predictive than a simple substitution profile which assigns a penalty for any point mutation.\\
The most interesting test performed here considers a realistic scenario where no information on the bound sequence is available. Even in this case MM/GBSA(+HB) energies are predictive.\\
This result has important consequences for the prediction of transcription factor binding sites which often use consensus methods. A prerequisite for the usefulness of consensus methods is that these are as independent of each other as possible. Since most methods use common prior knowledge and often related statistical methods, independence is not guaranteed. 
Methods which are based on completely independent principles, like those based on physical effective energy functions and free energy computations, offer a completely complementary methodology for deriving profile matrices for scanning entire genomes. The results reported here, with much caution because the structural model for the specific bound sequence is known and not modeled by homology or other methods, support usage of these methods for the identification of DNA-binding protein target sequences. In view of the very recent impressive results reported by the group of Baker \cite{Ashworth:2006} it is apparent that significant improvements to the approach described in this paper may be obtained by extensive refinement and screening of protein side chain conformation at protein-DNA interface.\\

\section*{Methods}
\vspace*{0.5cm}
\subsection*{Model building}

Atomic coordinates of the $\lambda$ repressor dimer bound to $O_L1$ DNA operator 
were taken 
from the 1.8 \AA~ resolution X-ray crystal structure deposited in the Protein Data 
Bank \cite{PDB:2000} (PDB code 1LMB). The operator is 17 base-pairs in length and is composed 
by two approximately symmetric parts, the "consensus half" (maintaining the notation of the PDB file, 
base-pairs A19-T23 to G11-C31) 
and the "non-consensus half" (base-pairs T3-A39 to G10-C32) (see Figure 7). 
Since the coordinates of the 
NH$_2$-terminal arm of the repressor bound to the non-consensus half operator 
were not available, the lacking amminoacids were added using the protein bound 
to the consensus half operator. Using the program ProFit V2.2 \cite{ProFit}, 

the C$_\alpha$ carbons of 
the proteins have been superimposed and afterward the amino acids of the rotated 
structure have been added to the other one. Since the detailed X-ray crystal 
structure is made up of $\lambda$ repressor dimer and $O_L1$ operator DNA while the 
experimental data concern the $O_R1$ site, the WHATIF \cite{Vriend:1990} program was used to 
substitute the base-pair at position 5 to obtain the wild-type $O_R1$ operator. 
All possible single base-pair substitutions within the DNA sequence were generated using the program WHATIF \cite{Vriend:1990}. 

\begin{figure}
\begin{center} 
\subfigure[]{\includegraphics[scale=0.5]{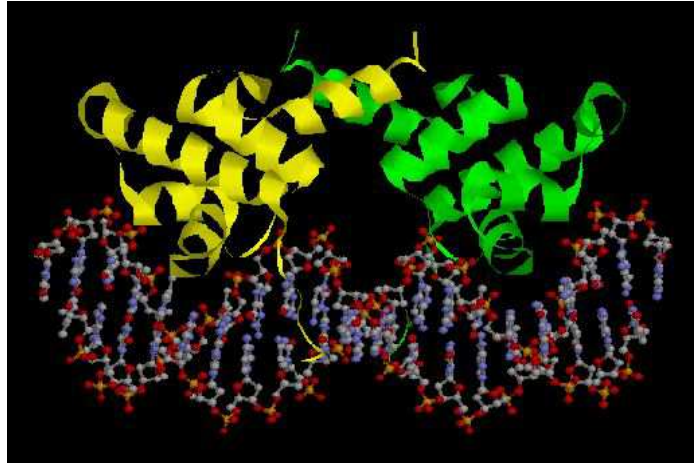}}
\subfigure[]{\includegraphics[scale=0.6]{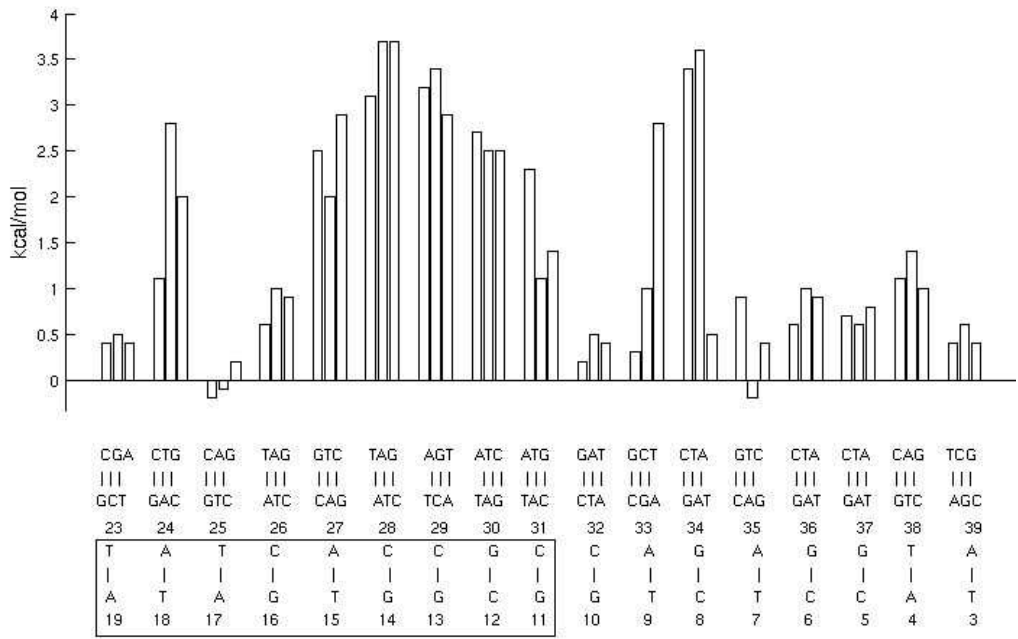}}
\caption{(a) Structure of the complex of $\lambda$ repressor \cite{Beamer:1992} with operator DNA. The protein
was crystallized with a 19-bp duplex of which the central 17 bps are shown. The consensus half is to the left.
(b) Relative free energy changes in the binding of $\lambda$ repressor to $O_R1$ on base substitutions. The figure shows
the change in affinity that results from each of the three possibile substitution at all 17 sites. The
left part represents the consensus half-operator (solid box) and the right half the non-consensus half-operator 
(redrawn from ref.\cite{Sarai:1989}).} 
\end{center}
\end{figure}

\subsection*{Molecular Dynamics Simulations}

Hydrogen atoms have been added using the program pdb2gmx of the GROMACS package \cite{GROMACS:1995,GROMACS:2001}. 
Every structure has been optimized performing 200 steps of energy 
minimization using the NAMD program, fixing all C$_\alpha$ carbons and 
phosphate groups 
coordinates. A dielectric constant of 10 has been employed with a cut-off of 12 
\AA~ for non-bonded interactions.\\
The net charge of the system 
($-36$) has been neutralized placing a corresponding number of sodium counterions 
in energetically favourable positions. 
The electrostatic potential was 
calculated via numerical solutions of the Poisson-Boltzmann equation using the 
University of Houston Brownian Dynamics (UHBD, version 6.x) program \cite{UHBD:1994,UHBD:1995}. A counterion was placed at the lowest potential position at 7.0 \AA~ from any heavy atom of the solute. The cycle was repeated until the net charge of the system was 0.\\
The complex and counterions were solvated in a box of TIP3P \cite{Jorgensen:1983}
water molecules using the solvate module in the program VMD \cite{Humphrey:1996}. The resulting system contained about 4200 solute atoms and 50400 solvent atoms. The 
coordinates of the solute were fixed and the solvent was energy minimized using 
100 steps of conjugate gradient. A solvent equilibration was carried out by 
performing molecular dynamics for 50 ps using a 1 fs time step to let the water 
molecules move to adjust to the conformation of the solute. The system was then 
energy minimized using 100 steps of conjugate gradient and, after 100 ps 
equilibration, 1-ns MD simulations was performed using a 2-fs timestep.
A snapshot of the trajectory was stored every 10 ps for later analysis. The 
shakeH algorithm was used in order to fix bond length 
between each hydrogen and its mother atom to its nominal value and to extend the simulation time-step \cite{Andersen:1983}. 
All molecular dynamics simulations of the complex were run under 
constant NPT conditions using the NAMD program \cite{Kale:1999}. 
The pressure of the system was coupled, through a 
Berendsen-thermostat \cite{Berendsen:1984}, to a pressure bath with target pressure 1.01325 bar and 
time constant 100 fs. The temperature has been kept to 300 K by simple velocity rescaling every picosecond. Long-range electrostatic interactions were 
treated by particle mesh Ewald (PME) method \cite{Darden:1993} employing a grid of 128x128x128 points. The cut-off was 12 \AA~ and the tolerance was $10^{-6}$ which resulted in an Ewald coefficient of 0.257952. The order for PME interpolation was 4.\\ 
The simulations were performed on 
a cluster composed by ten dual-processor nodes based on Intel XeonTM 2.8 GHz , 
with hyper-threading technology.

\subsection*{Free energy calculations}

The free energy of binding for each structure has been computed 
according to the framework reviewed by Gilson et al. \cite{Gilson:1997} who derived the expression of the free energy of binding in terms of the microscopic properties of the two 
molecules involved, using standard statistical thermodynamics. 
Here, similar to other works employing continuum methods several simplifications are adopted. 
The free energy 
of binding for each complex minus the entropic contribution is expressed as the sum of 
the interaction energy between the protein and the DNA $\Delta U(\vec{r_1} ,\dots,\vec{r_n})$ and a
solvation free energy term $\Delta G_{solv}$ :
\begin{equation}\label{eq1}
\Delta G = \Delta U(\vec{r_1} ,\dots,\vec{r_n}) + \Delta G_{solv}(\vec{r_1} ,\dots,\vec{r_n}) 
\end{equation}
It has been assumed that the entropy restriction in internal degrees of freedom and overall rotation and translation degrees of freedom is the same for all complexes.\\ 
The effect that association has on intramolecular energy has been neglected. Moreover no extended conformational search has been performed for protein side chains and DNA, partly because this task is not easily accomplished and partly because large conformational changes often result in large molecular mechanics energy changes, so we aimed at keeping the systems to be compared as close as possible.\\
The free energy of binding has been calculated using different methodologies detailed below. For all models alternative versions in which an energy term proportional to the number of hydrogen bonds has been added have been considered.\\
Except where noted, all contributions to the free energy of binding have been optimally scaled in order to best reproduce available experimental data (see later).\\

\subsubsection*{MM/DDDC-OONS}

In this method \cite{Oobatake:2003} electrostatic interactions have been estimated using a distance dependent dielectric constant (DDDC) while the solvation energy is proportional to the solvent accessible surface area through the atomic solvation parameters of Oobatake, Ooi, Nemethy and Scheraga (OONS) \cite{Oobatake:1993}.\\
All structures have been energy minimized with 200 conjugate gradient steps, using a distance-dependent dielectric constant (four values have been tested: $1r$, $2r$, $4r$, $8r$, with the distance r expressed in \AA) and a cut-off of 12 \AA.
The molecular mechanics interaction energy $U(\vec{r_1} ,\dots,\vec{r_n})$ was evaluated using CHARMM (version 27b2), a classic and well-tested molecular mechanics force-field \cite{CHARMM:1983,CHARMM:1998}. 
This term includes the nonbonded electrostatic and Van der Waals contributions. The solvation free energy term $G_{solv}$ has been calculated according to the model developed by Oobatake et al. \cite{Oobatake:2003}. This model consists in assigning every atom to one of 9 classes of chemical groups and assuming that the hydration 
free energy of every group $i$ in a solute is proportional to its solvent 
accessible surface area (SASA) $A_i$, because the group can directly interact only with water molecules at the surface.
\begin{equation}\label{eq2}
G_{solv} =\sum_{i}g_{i} ^{hyd} (A_i(complex) - A_i (protein) - A_i(DNA))
\end{equation}
The proportionality constants $g_{i} ^{hyd} $ have been determined from thermodynamic data on 
the transfer of small molecules from the gas phase into aqueous environment 
assuming the additivity of contributions from individual groups \cite{Oobatake:1993}.\\

\subsubsection*{MM/DDDC-HP}

In a very similar approach the OONS 9-parameter solvation model has been replaced by a simpler 2-parameter hydrophobic, polar (HP) solvation model.
Energy minimization protocol and tested values are the same as for the MM/DDDC-OONS for proper comparison.\\

\subsubsection*{MM/GBSA}

In this method the solvation 
free energy term is split in a polar (electrostatic) and a 
non-polar (hydrophobic) term.
\begin{equation}\label{eq3}
\Delta G_{solv}(\vec{r}) = \Delta G^{polar}(\vec{r}) + \Delta G^{non-polar}(\vec{r})
\end{equation}
The polar term is computed using the Generalized Born approach \cite{Qiu:1997}.
All complexes have been energy minimized by 200 conjugate gradients minimization steps using the generalized Born model as implemented in the CHARMM program, then the solute and solvation energy terms have been computed for both the complex and the isolated molecules. The binding energy was then computed by subtraction. Doubling the number of minimization steps does not 
affect significantly the results. \\
The non-polar term $G^{non-polar}$, which takes into account the tendency of the non-polar parts of 
the molecule to collapse, is taken to be proportional to the 
solvent-accessible surface area $A$, i.e. $G^{non-polar} = \gamma A$, where the surface 
tension coefficient $\gamma$ has been empirically determined to be equal to 20 cal \AA$^{-2}$ mol$^{-1}$ for this kind of applications \cite{Fogolari:2003}. \\
A variant of this methodology including splitting the solvent accessible surface area into a polar and a hydrophobic contribution (i.e. using two different surface tension coefficients), and including a term proportional to the number of hydrogen bonds has been considered here.\\

\subsection*{Finding optimal scaling factors}

The choice of methods and parameters in molecular mechanics/ implicit solvent methods is subject to large
uncertainties. In order to explore the best performance achievable with these methodologies, optimal scaling
factors for the different contributions were searched that could best reproduce the experimental data.
This approach is not new and it has been used successfully by other groups (see e. g. \cite{Morozov:2005}). In practice it is expected that proper scaling is able to compensate for the many inaccuracies of the model.\\
In general terms, the free energy of binding has been computed as a linear combination of contributions $E_i$, with corresponding coefficients $x_i$, i.e.:
\begin{equation}\label{eq4}
\Delta G_{binding} =  \sum _i x_i \Delta E_i 
\end{equation}
where $\Delta $ represents the difference between the complex and the isolated 
protein and DNA molecules.\\
Coefficients $x_1,\dots,x_n$ have been found in order to best reproduce the 52 experimentally available free energies of binding. 
Contributions have been arranged in a $52\times n$ matrix $A$ where each row 
corresponds to each structural model and each column corresponds to a different contribution to the free energy of binding. The experimental binding free energies have been arranged in a $52$-component vector $\Delta G_{exptl}$. 
The linear system $$Ax = \Delta G_{exptl}$$ where $x$ is the $n$-component vector of coefficients, has been solved (in a least square sense) using singular value decomposition \cite{Press:1995} and the best $x_i$ coefficients have been used to calculate binding energies $\Delta G_{calc}$.\\ 
A constant term takes into account the entropy loss upon complexation and other possible contributions identical for all complexes.\\
A linear model, compared to more sophisticated methods, has the advantage that the number of adjustable parameters is limited and easily interpretable in physical terms.\\
In the following we detail the contributions considered for each energetic model.\\
The free energy of binding has been computed for the MM/DDDC-OONS model according to the following equation:

\begin{eqnarray}\label{eq5}  
\Delta G_{MM/DDDC-OONS} & =  & x_{vdW} \Delta E_{vdW} +  x_{DDDC} \Delta E_{elec, DDDC} + x_{OONS} \Delta G_{OONS}  \nonumber \\
& & + x_{const.} (+ x_{HB} N_{HB})  
\end{eqnarray}
where $\Delta E_{vdW}$ is the van der Waal contribution, $\Delta E_{elec, DDDC}$ is the Coulombic energy, computed with a distance dependent dielectric constant, $\Delta E_{OONS}$ is the solvation energy according to the Oobatake et al. model \cite{Oobatake:1993} and $N_{HB}$ is the number of intermolecular hydrogen bonds.\\
As mentioned above, the coefficients bear physical meaning. For instance the term $x_{const}$ should account for rotational and translational entropy loss upon binding and it can be expected to be in the range 20-40 kcal/mol.\\
The term proportional to the number of hydrogen bonds was alternatively added 
in order to take into account possible inaccuracies in the treatment of these interactions by molecular mechanics and solvation terms. In practice every time this term is added the coefficients of molecular mechanics and solvation terms are greatly reduced thus avoiding double counting of hydrogen bond interactions.\\
A similar expression for the free energy of binding has been used for the MM/DDDC-HP model:
\begin{eqnarray}\label{eq6}
\Delta G_{MM/DDDC-HP}  & =  & x_{vdW} \Delta E_{vdW} +  x_{DDDC} \Delta E_{elec, DDDC} + x_{H} \Delta A_{H} +  x_{P} \Delta A_{P} \nonumber \\
& & + x_{const.}  (+ x_{HB} N_{HB})
\end{eqnarray}
Here the coefficients $x_{H}$ and $ x_{P}$ represent the surface tension coefficients multiplying hydrophobic and polar solvent accessible surface areas $\Delta A_{H}$ and $\Delta A_{P}$, respectively. We expect these coefficients to be in the range of tens of cal \AA$^{-2}$ mol$^{-1}$.\\
The solvent accessible area has been also splitted in polar and hydrophobic area for finding optimal scaling parameters for the MM/GBSA methodology:
\begin{eqnarray}\label{eq7}
\Delta G_{MM/GBSA} & = &  x_{vdW} \Delta E_{vdW} +  x_{Coul} \Delta E_{Coul} + x_{GB} \Delta G_{GB} + x_{H} \Delta A_{H} +  x_{P} \Delta A_{P} \nonumber\\
 & & + x_{const.} (+ x_{HB} N_{HB})
\end{eqnarray}
where $\Delta G_{GB}$ is the generalized Born solvation energy. The coefficients $x_{Coul}$ and $x_{GB}$ are exactly and roughly, respectively, inversely proportional to the effective dielectric constant and are thus expected to be in the range 0.05 to 1.0.\\

\subsection*{Possible pitfalls of the method}

Scaling energy terms for free energy evaluation of models which have been minimized without scaling such terms is clearly inconsistent. A correct procedure would be to iteratively find the optimal scaling factors, minimizing the energy using such scaling factors and repeating these two steps until convergence. 
This procedure faces some difficulties because an important term like the hydrogen bond term is discrete and does not have a counterpart in standard forcefields, where such interactions are described typically through electrostatic and van der Waals terms. Similarly the minimization of terms proportional to the solvent accessible surface area requires algorithms which are rarely available in molecular mechanics packages. 
A further difficulty is that any unbalance among forcefield terms might introduce distortions in molecular structure, notably of hydrogen bond lengths. 
Although the issue of iteratively fitting optimal scaling factors is worth being further investigated, here the approach of scaling factors has been applied in a more rough way. We have matched as far as possible the energetic model used for minimization with that used for fitting scaling factors, as mentioned above, but we have not minimized again the models using the scaling factors.
A similar mismatch between conformational sampling and energy evaluation is implicit in the analysis of molecular dynamics snapshots. 
Other sources of error in this case are the large conformational (and energetic) fluctuations molecules undergo during simulation and in general the inaccuracy of implicit solvent methods (used in energy evaluation) where small energy differences arise from subtraction of rather large values. It should be noted that for molecular dynamics snapshots inaccuracies do not cancel out because there are no restrained parts in the molecules.\\

\subsection*{DNA sequence dependent deformability}

An important aspect of protein-DNA interaction, addressed quantitatively
by Olson and co-workers \cite{Olson:1998}, is the capability of DNA
sequences to adopt specific
local conformations. The statistics of parameters and pairwise parameter
correlations
shows definite preferences. In the approach described above, changes in
intramolecular
energy terms are disregarded altogether by the assumption of rigid
docking.
The strains introduced in complex molecular structures, however, are
typically relaxed over the
structure and should have consequences on the intermolecular energy
terms too.
In order to assess the effect of DNA sequence dependent deformability we
followed the
approach of Olson and co-workers \cite{Olson:1998}, who made available
average parameters
for the six parameters describing local geometry of a base-pair step in
B-DNA, the force constant
parameters for all pairwise deviation from equilibrium values and a
program to analyse
DNA structures \cite{Lu:2003}. \\
The analysis was performed for the native structure parameters, simply
replacing the
identity of the base-pair mutated, and on the mutated structures,
minimized using the
generalized Born model.
For both cases poor correlation with experimental binding data was
found. Remarkably, however,
the native sequence was the third lowest energy sequence among all 52
sequences. Energy minimization
in general increases the energy associated with the deformability of
DNA.
Computation of the fitness of a sequence to local geometry parameters
gives important
informations although it is likely that the computed energy is not
accurate for conformations
far from equilibrium.
Inclusion of the DNA sequence dependent deformability energy in the
analyses detailed below
did not improve results significantly, notwithstanding the additional
scaling
parameter introduced for this purpose. For this reason this term was not
considered further.

\subsection*{Performance analysis} 

After fitting scaling factors to experimental data, the root mean square difference between calculated and experimental data was 
computed. This quantity can provide however a poor evaluation of the predictive power of the calculations when the test systems are very similar. Therefore the correlation coefficient between calculated and experimental data was also computed.
Optimal scaling factors were computed taking all the data available.\\

Fitting 52 experimental data with up to 7 parameters will 
always results in a positive correlation coefficient. In order to make sure
that the results obtained are significant we performed different kind of 
analyses:\\
i) a leave-one-out scheme has been adopted. All but one of the data were taken and the root mean square difference and correlation coefficient were computed using the set of data not used in the fitting procedure.\\
The same scheme has been applied to 5000 replicates with one third of the data 
left out of the fitting procedure and used for RMSD and correlation coefficient computation.\\
ii) the variance of each linear coefficient has been estimated from the 
multiple regression analysis using the variance/covariance matrix and the
square error of computed data, according to standard linear regression
procedures \cite{Berenson:1983}. In practice the standard deviation of experimental data has been
estimated as 
\begin{equation}\label{sigma1}
\sigma = \sqrt{\sum_i (G_{i, calc} - G_{i, exptl})^2 \over {n - m}}  
\end{equation}
Then the variance of each coefficient has been estimated from the variance/covariance matrix of coefficients:
\begin{equation}\label{sigma2}
\sigma^2 (x_j) = (A^T A)^{-1} _{jj}\sigma^2 
\end{equation}
The different models considered employ a different number of fitting parameters and therefore different performances are expected. Although these parameters are often correlated, the analysis of the variance gives an immediate clue as to which variables are more important.\\
iii) analysis of variance (ANOVA) calculations have been performed and a significance test based on the F-statistic and the corresponding confidence level has been computed \cite{Berenson:1983}. \\
iv) one thousand replicates of the original data has been generated with the 
column elements containing the experimental data randomly swapped. The average
of the correlation coefficient between swapped experimental data and fitted
data has been computed together with the standard deviation. The results of 
this computation (not reported) fully supports the results of the statistical 
analyses described above;\\
Finally, a useful alternative to assess the quality of $\Delta G_{calc}$ predictions and to compare the different models from a qualitative
point of view, consists in determining the number of ``correct predictions'', defined as the number of cases in which both 
$\Delta G_{exp}$ and $\Delta G_{calc}$ are  $<$1.0 kcal/mol, or $>$1.0 kcal/mol, or else separated by less than 0.3 kcal/mol. The threshold 
value of 1.0 kcal/mol requires some explanations.
 The experimental values of the free energy change relative to the wild-type operator $O_R1$ have been calculated using the 
equation $\Delta G_{exp}$ = - 0.546 ln ($K_d$ of substituted sequence)/($K_d$ of $O_R1$)
 after having determined the dissociation constant of every mutant. It is simple to verify that the threshold value of 1.0 
kcal/mol corresponds to a remarkable reduction in the dissociation constant of the mutant (ca. 5-fold), with respect to the 
dissociation constant of the wild-type operator ($K_d$ of $O_R1$ = $10^{-9}$), whereas values of $\Delta G$ higher than 1.0 kcal/mol 
correspond to a reduction in the dissociation constant from 5 ($\Delta G_{exp}$ = 1.0 kcal/mol) to 25-fold  ($\Delta G_{exp}$=3.4, which is the maximum value of $\Delta G_{exp}$). 
Therefore it is reasonable to define $\Delta G_{calc}$ as correct, if both $\Delta G_{exp}$ and $\Delta G_{calc}$ are 
in one of the defined intervals or even if the difference $D=|\Delta G_{exp}-\Delta G_{calc}|$ is lower than 0.3 kcal/mol, which 
corresponds to a ratio between the dissociation constant of a mutant and the dissociation constant of the wild-type complex lower
 than 2.0.\\ 

\subsection*{Analysis of non-specific protein-DNA binding}

One thousand random DNA sequences were generated and
the corresponding structural models were generated by performing
mutations on the double stranded DNA in the complex crystal structure
using the program WHATIF \cite{Vriend:1990}.
The resulting dataset of complexes was assumed to be representative of
non-specific protein-DNA complexes.
We are interested in understanding how reliable is the method for predicting putative binding sites. The so-called Z-score of the specific bound sequence compared to random sequences has been considered.
The Z-score is defined here as the distance of the free energy computed for the specific bound 17-mer ($\Delta G$) from the average non-specific binding energy ($<\Delta G>$), normalized by the standard deviation of the computed non-specific binding energies ($\sigma_G$).
\begin{equation}
Z-score = {{\Delta G - <\Delta G>}\over {\sigma_G}}
\end{equation}
Averages are performed over the one thousand random sequences.
A large Z-score implies that the specific bound sequence can be distinguished from other non-specific bound sequences.
The structures were energy minimized using the same protocol used for MM/GBSA
free energy estimation.
For all minimized complexes the Coulombic energy, van der Waals energy, GB solvation energy, polar and apolar surface accessible area and intermolecular hydrogen bonds number were tabulated.
For each model $i$ of the 1000 random DNA sequence complexes the binding energy $G(i)$ has been computed using different amounts of the
experimental information available.
Different analyses, detailed in the Results section, were performed.\\ 
The possibility of using the data computed on the set of non-specific complexes for defining a profile of the recognized DNA sequences has been explored as follows.
The calculated binding energy values for the set of non-specific complexes were summarised in a set of 68 values corresponding to the average contribution to the binding free energy of each possible of the 4 bases at each of the possible 17 bound sequence positions.  
These 68 values have been derived as follows.
Possible substitutions are indexed from 1 to 4 for A, C, G and T, respectively.
A $1000 \times 68$ matrix $A$ was set where each element
$A(i,j)$ is 1.0 or 0.0 if the base at position $j/4$ (rounded at the closer upper integer) has index $j \bmod 4 $ in sequence i.
The set of $68$ substitution free energies $x(j)$ were found by solving (in a root mean square error sense) the overdetermined equation $Ax = G$.
The resulting 68-element vector $x$ was arranged in a $17 \times 4$ matrix.
Variants on this procedure are described in the Results section according
to the level of information available included in the analysis.\\

\subsection*{Scanning of bacteriophage $\lambda$ genome}
The free energy matrix derived from the analysis of non-specific protein-DNA complexes was used to score all 17-mer subsequences in the bacteriophage $\lambda$ genome (Accession number: NC\_001416.1, 48502 base-pairs) on both strands. In principle the score represents the free energy of binding of the 17-mer considered.\\ 
Reference "experimental" binding free energy values, for comparison with computed data, were obtained under the hypothesis of additivity \cite{Benos:2000,Benos:2002} using experimental data on single base-pair mutants.\\

\section*{Authors' contributions}
EM designed and performed most tests and analyses,
and wrote part of the code used for the analyses.
MC and FF conceived the project,
supervised the work and designed some of the analyses.
FF wrote part of the code used for the analyses.
All authors have read and approved the final version of the manuscript.

\section*{Acknowledgments}
We wish to thank Drs. G. Tecchiolli and P. Zuccato of Exadron, the HPC
Division of the Eurotech Group, for providing hardware and expert technical
assistance.\\
Dr. M. Isola of the University of Udine is gratefully acknowledged
for helpful discussions on statistical aspects of multiple regressions.\\
EM wishes to thank Profs. C. Destri, G. Marchesini and F. Rapuano for helpful discussions.\\
Part of the research was funded by FIRB grant RBNE03B8KK from the Italian Ministry for Education, University and Research.\\

\newpage
\section*{Tables}

\subsection*{Table 1 - Optimal scaling factors for the MM/DDDC-OONS model and the MM/DDDC-HP model }
\linespread{0.9}
\begin{table}[!h]
\begin{center} 
{\small
\begin{tabular}[t]{cccccc}
\hline
\multicolumn{2}{c}{MM/DDDC-OONS}  &  \bf 1r & \bf 2r  & \bf 4r  &  \bf 8r \\
\hline
\multicolumn{2}{c}{\bf $ x_{vdW}$ } & 0.075 (0.071) & 0.041 (0.100) & 0.043 (0.111) & 0.025 (0.116) \\
\multicolumn{2}{c}{\bf $ x_{DDDC}$ } & 0.083 (0.019) & 0.184 (0.037) & 0.359 (0.076) & 0.802 (0.159) \\
\multicolumn{2}{c}{\bf $ x_{OONS}$ } & 0.072 (0.070) & -0.020 (0.075) & -0.084 (0.061) & -0.043 (0.076) \\
\multicolumn{2}{c}{\bf $ x_{const}$ } & 65.254 (20.806) & 73.523 (21.615) & 75.782 (23.697) & 64.715 (26.116) \\
\hline
\multicolumn{2}{c}{ MM/DDDC-OONS (+HB) }  &  \bf 1r & \bf 2r  & \bf 4r  &  \bf 8r \\
\hline
\multicolumn{2}{c}{\bf $ x_{vdW}$ } & 0.075 (0.072) & 0.040 (0.100) & 0.042 (0.109) & 0.034 (0.113) \\
\multicolumn{2}{c}{\bf $ x_{DDDC}$ } & 0.068 (0.030) & 0.154 (0.046) & 0.286 (0.089) & 0.637 (0.179) \\
\multicolumn{2}{c}{\bf $ x_{OONS}$ } & 0.075 (0.071) & -0.019 (0.075) & -0.066 (0.061) & -0.010 (0.076) \\
\multicolumn{2}{c}{\bf $ x_{HB}$ } & -0.151 (0.249) & -0.166 (0.152) & -0.226 (0.145) & -0.269 (0.144) \\
\multicolumn{2}{c}{\bf $ x_{const}$ } & 57.867 (24.223) & 68.819 (21.992) & 69.397 (23.718) & 58.463 (25.680) \\
\hline
\multicolumn{2}{c}{MM/DDDC-HP} & \bf 1r & \bf 2r & \bf 4r & \bf 8r    \\
\hline
\multicolumn{2}{c}{\bf $ x_{vdW}$ } & 0.144 (0.065) & 0.215 (0.106) & 0.085 (0.115) & 0.133 (0.128) \\
\multicolumn{2}{c}{\bf $ x_{DDDC}$ } & 0.076 (0.017) & 0.221 (0.033) & 0.402 (0.075) & 0.844 (0.153) \\
\multicolumn{2}{c}{\bf $ x_{P}$ } & -0.018 (0.008) & -0.027 (0.011) & -0.007 (0.011) & -0.012 (0.012) \\
\multicolumn{2}{c}{\bf $ x_{H}$ } & -0.028 (0.007) & -0.021 (0.007) & -0.013 (0.008) & -0.015 (0.008) \\
\multicolumn{2}{c}{\bf $ x_{const}$ } & -4.023 (26.223) & 21.075 (24.822) & 38.294 (32.091) & 31.927 (29.069) \\
\hline
\multicolumn{2}{c}{ MM/DDDC-HP + (HB)} & \bf 1r & \bf 2r & \bf 4r & \bf 8r    \\
\hline
\multicolumn{2}{c}{\bf $ x_{vdW}$ } & 0.144 (0.065) & 0.221 (0.104) & 0.105 (0.111) & 0.175 (0.123) \\
\multicolumn{2}{c}{\bf $ x_{DDDC}$ } & 0.059 (0.027) & 0.185 (0.040) & 0.281 (0.089) & 0.623 (0.172) \\
\multicolumn{2}{c}{\bf $ x_{P}$ } & -0.018 (0.008) & -0.029 (0.011) & -0.011 (0.011) & -0.018 (0.011) \\
\multicolumn{2}{c}{\bf $ x_{H}$ } & -0.029 (0.007) & -0.022 (0.007) & -0.017 (0.008) & -0.017 (0.007) \\
\multicolumn{2}{c}{\bf $ x_{HB}$ } & -0.181 (0.218) & -0.210 (0.138) & -0.327 (0.143) & -0.332 (0.137) \\
\multicolumn{2}{c}{\bf $ x_{const}$ } & -13.355 (28.624) & 12.784 (25.076) & 21.219 (31.635) & 20.937 (28.041) \\
\hline
\end{tabular}
\caption{Optimal scaling factors for the MM/DDDC-OONS model and the MM/DDDC-HP model. Standard deviations (see Methods section) are given in parentheses.
}
\label{scale_dddc}}
\end{center}
\end{table}

\newpage
\subsection*{Table 2 - RMSD and correlation coefficients}
\linespread{0.9}
\begin{table}[!h]
\begin{center}
{\small
\begin{tabular}{ccccccccc}
\hline
 &  & & \multicolumn{2}{c}{\bf all} & &  \multicolumn{2}{c}{\bf loo}  & \\
\multicolumn{2}{c}{} & \multicolumn{1}{c}{\bf rmsd} &  \multicolumn{1}{c}{\bf r} & &  &   \multicolumn{1}{c}{\bf rmsd} & \multicolumn{1}{c}{\bf r} \\
\hline
\multicolumn{2}{c}{\bf MM/DDDC-OONS } &   & & \multicolumn{1}{c}{\bf F(3,48)} &  {\bf p} &  &  & \\
\hline
\multicolumn{2}{c}{\bf 1r}  & 0.990 &  0.538 & 6.527 & $<$ 0.001 & 1.086 & 0.406 \\
\multicolumn{2}{c}{\bf 2r}  & 0.886 &  0.656 & 12.108 & $<$ 0.001 & 0.967 & 0.575 \\
\multicolumn{2}{c}{\bf 4r}  & 0.857 &  0.684 & 14.081 & $<$ 0.001 & 0.926 & 0.619 \\
\multicolumn{2}{c}{\bf 8r}  & 0.969 &  0.673 & 13.253 & $<$ 0.001 & 0.943 & 0.600 \\
\hline
\multicolumn{2}{c}{\bf MM/DDDC-OONS (+HB) } &   & & \multicolumn{1}{c}{\bf F(4,47)} &  {\bf p} &  &  & \\
\hline
\multicolumn{2}{c}{\bf 1r}  & 0.986 &  0.543 & 4.923 &  0.002 & 1.109 & 0.375 \\
\multicolumn{2}{c}{\bf 2r}  & 0.875 &  0.667 &  9.420 & $<$ 0.001 & 0.967 & 0.576 \\
\multicolumn{2}{c}{\bf 4r}  & 0.836 &  0.703 & 11.471 & $<$ 0.001 & 0.918 & 0.629 \\
\multicolumn{2}{c}{\bf 8r}  & 0.838 &  0.701 & 11.332 & $<$ 0.001 & 0.922 & 0.624 \\
\hline
\multicolumn{2}{c}{\bf MM/DDDC-HP } &   & & \multicolumn{1}{c}{\bf F(4,47)} &  {\bf p} &  &  & \\
\hline
\multicolumn{2}{c}{\bf 1r}  & 0.863 &  0.678 & 10.007 & $<$ 0.001 & 0.954 & 0.591 \\
\multicolumn{2}{c}{\bf 2r}  & 0.803 &  0.730 & 13.416 & $<$ 0.001 & 0.890 & 0.658 \\
\multicolumn{2}{c}{\bf 4r}  & 0.850 &  0.690 & 10.695 & $<$ 0.001 & 0.944 & 0.601 \\
\multicolumn{2}{c}{\bf 8r}  & 0.840 &  0.699 & 11.243 & $<$ 0.001 & 0.933 & 0.614 \\
\hline
\multicolumn{2}{c}{\bf MM/DDDC-HP (+HB) } &   & & \multicolumn{1}{c}{\bf F(5,46)} &  {\bf p} &  &  & \\
\hline
\multicolumn{2}{c}{\bf 1r}  & 0.857 &  0.684 & 8.089 &  $<$ 0.001 & 0.967 & 0.578 \\
\multicolumn{2}{c}{\bf 2r}  & 0.783 &  0.745 & 11.500 & $<$ 0.001 & 0.888 & 0.662 \\
\multicolumn{2}{c}{\bf 4r}  & 0.805 &  0.728 & 10.369 & $<$ 0.001 & 0.903 & 0.645 \\
\multicolumn{2}{c}{\bf 8r}  & 0.791 &  0.739 & 11.099 & $<$ 0.001 & 0.889 & 0.659 \\
\hline
\multicolumn{2}{c}{\bf MM/GBSA } &   & & \multicolumn{1}{c}{\bf F(5,46)} &  {\bf p} &  &  & \\
\hline
\multicolumn{2}{c}{}  & 0.992 &  0.664 & 7.258 & $<$ 0.001  & 1.109 & 0.551 \\
\hline
\multicolumn{2}{c}{\bf MM/GBSA (+HB) } &   & & \multicolumn{1}{c}{\bf F(6,45)} &  {\bf p} &  &  & \\
\hline
\multicolumn{2}{c}{}  & 0.782 &  0.746 & 9.413 & $<$ 0.001  & 0.928 & 0.630 \\
\hline
\end{tabular}
\caption{RMSD and correlation coefficients (r) between calculated and
experimental values using all available data (all) and the leave-one-out
cross validation technique (loo).  }
\label{rmsd_cc}
}
\end{center}
\end{table}

\newpage

\subsection*{ Table 3 - Components of the free energies (MM/DDDC-HP(+HB) model)}
\linespread{0.9}
\begin{table}[!h]
\begin{center}
{\scriptsize
\begin{tabular}[h]{ccccccccc}
\hline
{\bf Original base} & {\bf Mutated} &{$\bf \Delta G_{exp}$}&{$\bf \Delta G_{calc}$} &{\bf vdW} & {\bf Coul} & {\bf H}& {\bf P} & {\bf HB}\\
\hline 
 T3-A39   & C-G &   0.400 &    0.311 &  -50.460 &  -12.925 & 36.028 & 47.532 &  -6.508  \\ 
          & G-C &   0.600 &    0.204 &  -50.326 &  -12.684 & 35.875 & 47.203 &  -6.508  \\ 
          & A-T &   0.400 &   -0.091 &  -50.384 &  -12.669 & 35.965 & 46.861 &  -6.508  \\ 
 A4-T38   & C-G &   1.000 &    1.245 &  -49.982 &  -13.112 & 35.965 & 48.057 &  -6.327  \\ 
          & T-A &   1.400 &    2.666 &  -49.455 &  -13.227 & 35.992 & 48.858 &  -6.147  \\ 
          & G-C &   1.100 &    1.456 &  -49.704 &  -13.341 & 36.114 & 48.070 &  -6.327  \\ 
 C5-G37   & T-A &   0.800 &    1.681 &  -50.791 &  -13.042 & 36.270 & 49.108 &  -6.508  \\ 
          & A-T &   0.600 &    0.404 &  -50.786 &  -12.985 & 36.360 & 47.860 &  -6.689  \\ 
          & G-C &   0.700 &    0.659 &  -50.921 &  -12.952 & 36.417 & 47.978 &  -6.508  \\ 
 C6-G36   & T-A &   0.900 &    1.176 &  -50.637 &  -12.833 & 36.068 & 48.622 &  -6.689  \\ 
          & A-T &   1.000 &    1.918 &  -50.517 &  -12.560 & 35.698 & 49.161 &  -6.508  \\ 
          & G-C &   0.600 &    1.016 &  -50.216 &  -12.680 & 35.837 & 47.939 &  -6.508  \\ 
 T7-A35   & G-C &   0.400 &    0.018 &  -50.525 &  -13.005 & 35.845 & 47.387 &  -6.327  \\ 
          & A-T &  -0.200 &    0.301 &  -50.387 &  -13.076 & 35.731 & 47.715 &  -6.327  \\ 
          & C-G &   0.900 &    0.725 &  -50.472 &  -12.987 & 36.267 & 47.781 &  -6.508  \\ 
 C8-G34   & T-A &   0.500 &    1.878 &  -50.138 &  -13.246 & 36.020 & 48.924 &  -6.327  \\ 
          & A-T &   3.600 &    2.417 &  -49.584 &  -12.945 & 35.415 & 49.213 &  -6.327  \\ 
          & G-C &   3.400 &    1.965 &  -49.762 &  -13.259 & 36.086 & 48.583 &  -6.327  \\ 
 T9-A33   & A-T &   2.800 &    2.387 &  -49.570 &  -12.933 & 36.187 & 48.386 &  -6.327  \\ 
          & G-C &   1.000 &    1.111 &  -49.390 &  -13.552 & 36.308 & 47.427 &  -6.327  \\ 
          & C-G &   0.300 &    0.489 &  -50.729 &  -12.988 & 36.235 & 47.834 &  -6.508  \\ 
G10-C32   & A-T &   0.400 &    0.974 &  -50.711 &  -12.531 & 36.073 & 48.187 &  -6.689  \\ 
          & T-A &   0.500 &    1.377 &  -50.195 &  -13.338 & 36.243 & 48.530 &  -6.508  \\ 
          & C-G &   0.200 &    0.654 &  -51.175 &  -12.775 & 36.602 & 48.227 &  -6.870  \\ 
G11-C31   & C-G &   1.400 &   -0.061 &  -50.889 &  -13.098 & 36.203 & 47.768 &  -6.689  \\ 
          & A-T &   1.100 &    0.760 &  -50.512 &  -13.375 & 35.925 & 48.766 &  -6.689  \\ 
          & T-A &   2.300 &    0.451 &  -51.217 &  -12.921 & 36.481 & 48.332 &  -6.870  \\ 
C12-G30   & G-C &   2.500 &    2.764 &  -48.826 &  -13.604 & 36.091 & 48.425 &  -5.966  \\ 
          & A-T &   2.500 &    1.043 &  -50.496 &  -12.687 & 36.059 & 48.030 &  -6.508  \\ 
          & T-A &   2.700 &    2.161 &  -49.468 &  -13.250 & 35.898 & 48.845 &  -6.508  \\ 
G13-C29   & A-T &   2.900 &    2.337 &  -50.000 &  -12.972 & 36.226 & 48.766 &  -6.327  \\ 
          & C-G &   3.400 &    2.413 &  -49.823 &  -12.729 & 36.460 & 48.188 &  -6.327  \\ 
          & T-A &   3.200 &    2.546 &  -49.982 &  -12.803 & 36.574 & 48.621 &  -6.508  \\ 
G14-G28   & C-G &   3.700 &    2.728 &  -49.462 &  -12.929 & 36.226 & 48.215 &  -5.966  \\ 
          & T-A &   3.700 &    2.834 &  -49.692 &  -12.379 & 35.851 & 48.556 &  -6.147  \\ 
          & A-T &   3.100 &    2.202 &  -50.081 &  -13.153 & 36.145 & 48.793 &  -6.147  \\ 
T15-A27   & G-C &   2.900 &    1.728 &  -50.522 &  -12.607 & 36.887 & 47.834 &  -6.508  \\ 
          & A-T &   2.000 &    1.089 &  -50.515 &  -12.812 & 36.617 & 47.663 &  -6.508  \\ 
          & C-G &   2.500 &    0.966 &  -50.540 &  -12.811 & 36.544 & 47.637 &  -6.508  \\ 
G16-C26   & C-G &   0.900 &    1.793 &  -49.675 &  -12.665 & 36.140 & 47.676 &  -6.327  \\ 
          & T-A &   1.000 &    1.767 &  -50.112 &  -12.163 & 35.993 & 47.912 &  -6.508  \\ 
          & A-T &   0.600 &    1.276 &  -50.287 &  -13.133 & 36.279 & 48.280 &  -6.508  \\ 
A17-T25   & C-G &   0.200 &    1.168 &  -50.790 &  -12.598 & 36.796 & 47.624 &  -6.508  \\ 
          & T-A &  -0.100 &    1.134 &  -50.710 &  -12.660 & 36.638 & 47.729 &  -6.508  \\ 
          & G-C &  -0.200 &    1.256 &  -50.358 &  -12.788 & 36.760 & 47.506 &  -6.508  \\ 
T18-A24   & C-G &   2.000 &    1.796 &  -49.628 &  -13.518 & 36.648 & 47.978 &  -6.327  \\ 
          & A-T &   2.800 &    2.799 &  -49.290 &  -13.722 & 36.428 & 48.885 &  -6.147  \\ 
          & G-C &   1.100 &    1.749 &  -49.774 &  -13.455 & 36.261 & 48.399 &  -6.327  \\ 
A19-T23   & T-A &   0.400 &    1.239 &  -50.111 &  -12.649 & 36.212 & 47.650 &  -6.508  \\ 
          & C-G &   0.500 &    1.378 &  -50.196 &  -12.587 & 36.216 & 47.808 &  -6.508  \\ 
          & G-C &   0.400 &    0.780 &  -50.410 &  -12.892 & 36.243 & 47.703 &  -6.508  \\ 
wild-type & --- &   0.000 &    1.158 &  -50.404 &  -13.063 & 36.234 & 48.254 &  -6.508  \\ 
\hline
\end{tabular}}
\caption{Components of the free energies (in kcal/mol) calculated according to the MM/DDDC-HP(+HB) model. The distance dependent 
dielectric constant is 1r. The constant term $x_{const}$ is -13.355 kcal/mol }
\label{dddc_hp_all}
\end{center}
\end{table}

\newpage

\subsection*{Table 4 - Optimal scaling factors (MM/GBSA model).}
\linespread{0.9}
\begin{table}[!h]
\begin{center} 
\begin{tabular}[t]{ccccc}
\hline
\multicolumn{2}{c}{\bf MM/GBSA }      &   \bf $\lambda$   &  \bf Cro & \bf $\lambda$ + Cro   \\
\hline
\multicolumn{2}{c}{\bf $ x_{Coul}$ }  &  0.157 (0.045)    &  0.132 (0.067)   &  0.153 (0.037)    \\
\multicolumn{2}{c}{\bf $ x_{vdW}$ }   &  0.252 (0.124)    &  0.332 (0.132)   &  0.313 (0.086)    \\
\multicolumn{2}{c}{\bf $ x_{GB}$ }    &  0.142 (0.046)    &  0.121 (0.069)   &  0.145 (0.038)    \\
\multicolumn{2}{c}{\bf $ x_{P}$ }     & -0.010 (0.012)    &  0.019 (0.012)   & -0.009 (0.006)    \\
\multicolumn{2}{c}{\bf $ x_{H}$ }     & -0.029 (0.008)    & -0.009 (0.011)   & -0.031 (0.005)    \\
\multicolumn{2}{c}{\bf $ x_{const}$ } & -11.744 (28.025)  & 70.791 (37.615)  & -19.519(11.903)   \\
\hline
\multicolumn{2}{c}{\bf MM/GBSA(+HB)}  &  \bf $\lambda$    &  \bf Cro & \bf $\lambda$ + Cro        \\
\hline
\multicolumn{2}{c}{\bf $ x_{Coul}$ }  &  0.042 (0.053)    &  0.016  (0.066)  &  0.051  (0.041)    \\
\multicolumn{2}{c}{\bf $ x_{vdW}$ }   &  0.206 (0.113)    &  0.219  (0.118)  &  0.262  (0.079)    \\
\multicolumn{2}{c}{\bf $ x_{GB}$ }    &  0.025 (0.054)    &  0.002  (0.068)  &  0.043  (0.041)    \\
\multicolumn{2}{c}{\bf $ x_{P}$ }     & -0.016 (0.011)    & -0.001  (0.012)  & -0.024  (0.006)    \\
\multicolumn{2}{c}{\bf $ x_{H}$ }     & -0.023 (0.008)    & -0.005  (0.010)  & -0.029  (0.005)    \\
\multicolumn{2}{c}{\bf $ x_{HB}$ }    & -0.533 (0.156)    & -0.879  (0.224)  & -0.561  (0.125)    \\
\multicolumn{2}{c}{\bf $ x_{const}$ } & 13.699 (26.308)   & 65.011 (32.867)  & -21.595 (10.908)   \\
\hline

\end{tabular}
\caption{Optimal scaling factors for the MM/GBSA model and the MM/GBSA(+HB) model: analysis of the $\lambda$ repressor-operator system ($\lambda$), of the $Cro$-operator system ($Cro$) and the joint analysis ($\lambda + Cro$) . Standard deviations (see Methods section) are given in parentheses.
}
\label{scale_gb_pb_sa}
\end{center}
\end{table}

\newpage
\subsection*{Table 5 - Components of the free energies (MM/GBSA model) }
\linespread{0.9}
\begin{table}[!h]
\begin{center}
{\scriptsize
\begin{tabular}[h]{cccccccccc}
\hline
{\bf Original base} & {\bf Mutated} &{$\bf \Delta G_{exp}$}&{$\bf \Delta G_{calc}$} &{\bf Coul} & {\bf vdW} & {\bf GB}& {\bf P} & {\bf H} & {\bf HB}\\
\hline 
  T3-A39  & C-G &   0.400 &   0.115 & -79.691 & -33.547 &  50.408 &  30.607 &  37.296 & -18.658 \\ 
          & G-C &   0.600 &   0.525 & -79.292 & -33.467 &  50.226 &  30.690 &  37.327 & -18.658 \\ 
          & A-T &   0.400 &   1.165 & -78.394 & -33.432 &  49.648 &  30.588 &  37.181 & -18.125 \\ 
  A4-T38  & C-G &   1.000 &   0.777 & -81.129 & -33.803 &  51.302 &  30.367 &  37.934 & -17.592 \\ 
          & T-A &   1.400 &   1.036 & -81.246 & -33.723 &  51.482 &  30.201 &  38.215 & -17.592 \\ 
          & G-C &   1.100 &   1.613 & -79.949 & -33.852 &  50.681 &  30.609 &  38.017 & -17.592 \\ 
  C5-G37  & T-A &   0.800 &   2.136 & -80.893 & -33.864 &  51.109 &  30.552 &  39.125 & -17.592 \\ 
          & A-T &   0.600 &   0.996 & -79.955 & -33.842 &  50.645 &  30.484 &  38.090 & -18.125 \\ 
          & G-C &   0.700 &   0.377 & -81.302 & -33.810 &  51.382 &  30.589 &  37.944 & -18.125 \\ 
  C6-G36  & T-A &   0.900 &   1.362 & -80.476 & -33.730 &  50.877 &  30.661 &  38.456 & -18.125 \\ 
          & A-T &   1.000 &   0.622 & -79.641 & -33.667 &  50.411 &  30.314 &  38.163 & -18.658 \\ 
          & G-C &   0.600 &   1.110 & -80.186 & -33.579 &  50.678 &  30.491 &  38.132 & -18.125 \\ 
  T7-A35  & G-C &   0.400 &  -0.082 & -80.781 & -33.792 &  51.084 &  30.579 &  37.254 & -18.125 \\ 
          & A-T &  -0.200 &   0.318 & -80.382 & -33.960 &  50.847 &  30.200 &  38.038 & -18.125 \\ 
          & C-G &   0.900 &   1.146 & -80.387 & -33.817 &  50.873 &  30.729 &  37.641 & -17.592 \\ 
  C8-G34  & T-A &   0.500 &   1.350 & -80.849 & -33.862 &  51.113 &  30.593 &  38.247 & -17.592 \\ 
          & A-T &   3.600 &   3.360 & -77.402 & -33.305 &  49.165 &  30.110 &  38.152 & -17.059 \\ 
          & G-C &   3.400 &   2.435 & -78.703 & -33.834 &  49.920 &  30.582 &  37.829 & -17.059 \\ 
  T9-A33  & A-T &   2.800 &   2.732 & -78.808 & -33.818 &  50.131 &  30.297 &  38.289 & -17.059 \\ 
          & G-C &   1.000 &   1.489 & -79.210 & -34.213 &  50.394 &  30.812 &  37.599 & -17.592 \\ 
          & C-G &   0.300 &   0.714 & -80.866 & -33.839 &  51.216 &  30.778 &  37.851 & -18.125 \\ 
 G10-C32  & A-T &   0.400 &   1.868 & -79.605 & -33.008 &  50.347 &  30.450 &  38.110 & -18.125 \\ 
          & T-A &   0.500 &   1.034 & -80.384 & -34.230 &  50.938 &  30.722 &  38.414 & -18.125 \\ 
          & C-G &   0.200 &  -0.182 & -81.436 & -34.074 &  51.523 &  30.789 &  37.975 & -18.658 \\ 
 G11-C31  & C-G &   1.400 &   1.533 & -78.037 & -33.806 &  49.387 &  30.408 &  38.007 & -18.125 \\ 
          & A-T &   1.100 &   1.009 & -79.919 & -33.894 &  50.529 &  30.555 &  38.163 & -18.125 \\ 
          & T-A &   2.300 &   0.731 & -78.332 & -33.617 &  49.511 &  30.299 &  38.362 & -19.191 \\ 
 C12-G30  & G-C &   2.500 &   3.881 & -78.666 & -33.416 &  50.051 &  30.575 &  38.163 & -16.526 \\ 
          & A-T &   2.500 &   2.061 & -78.455 & -33.360 &  49.873 &  30.306 &  37.588 & -17.592 \\ 
          & T-A &   2.700 &   2.656 & -78.889 & -33.693 &  50.116 &  30.223 &  38.257 & -17.059 \\ 
 G13-C29  & A-T &   2.900 &   1.744 & -79.600 & -33.700 &  50.381 &  30.624 &  37.933 & -17.592 \\ 
          & C-G &   3.400 &   1.941 & -80.240 & -33.646 &  50.825 &  30.805 &  38.090 & -17.592 \\ 
          & T-A &   3.200 &   2.131 & -79.467 & -33.720 &  50.343 &  30.737 &  38.665 & -18.125 \\ 
 G14-G28  & C-G &   3.700 &   3.688 & -78.059 & -33.822 &  49.562 &  30.408 &  38.425 & -16.526 \\ 
          & T-A &   3.700 &   2.495 & -78.334 & -33.204 &  49.700 &  30.231 &  37.996 & -17.592 \\ 
          & A-T &   3.100 &   1.703 & -80.426 & -33.916 &  50.878 &  30.593 &  38.467 & -17.592 \\ 
 T15-A27  & G-C &   2.900 &   1.755 & -79.475 & -33.371 &  50.375 &  30.884 &  37.766 & -18.125 \\ 
          & A-T &   2.000 &   1.946 & -79.417 & -33.359 &  50.311 &  30.820 &  37.484 & -17.592 \\ 
          & C-G &   2.500 &   1.088 & -80.371 & -33.422 &  50.848 &  30.775 &  37.683 & -18.125 \\ 
 G16-C26  & C-G &   0.900 &   1.507 & -79.347 & -33.801 &  50.310 &  30.514 &  37.724 & -17.592 \\ 
          & T-A &   1.000 &   1.411 & -79.452 & -33.016 &  50.268 &  30.302 &  37.735 & -18.125 \\ 
          & A-T &   0.600 &   1.490 & -79.733 & -33.915 &  50.492 &  30.521 &  38.550 & -18.125 \\ 
 A17-T25  & C-G &   0.200 &   0.100 & -81.783 & -33.375 &  51.717 &  30.952 &  37.014 & -18.125 \\ 
          & T-A &  -0.100 &   0.258 & -81.595 & -33.370 &  51.621 &  30.858 &  37.171 & -18.125 \\ 
          & G-C &  -0.200 &   0.741 & -79.839 & -33.332 &  50.528 &  30.608 &  37.202 & -18.125 \\ 
 T18-A24  & C-G &   2.000 &   1.549 & -81.158 & -33.893 &  51.430 &  30.971 &  38.090 & -17.592 \\ 
          & A-T &   2.800 &   1.747 & -81.528 & -33.712 &  51.592 &  30.487 &  37.735 & -16.526 \\ 
          & G-C &   1.100 &   1.382 & -80.583 & -33.977 &  51.012 &  30.627 &  38.195 & -17.592 \\ 
 A19-T23  & T-A &   0.400 &   1.602 & -78.329 & -33.474 &  49.625 &  30.680 &  37.526 & -18.125 \\ 
          & C-G &   0.500 &   0.917 & -79.871 & -33.472 &  50.549 &  30.684 &  37.453 & -18.125 \\ 
          & G-C &   0.400 &   0.878 & -80.353 & -33.573 &  50.784 &  30.669 &  37.777 & -18.125 \\ 
wild-type & --- &   0.000 &   1.120 & -79.937 & -33.811 &  50.610 &  30.593 &  38.090 & -18.125 \\ 
\hline
\end{tabular}}
\caption{Components of the free energies (in kcal/mol) calculated according to the MM/GBSA model. 
The constant term $x_{const}$ is 13.699 kcal/mol }
\label{gb_hp_all}
\end{center}
\end{table}

\newpage

\subsection*{Table 6 - RMSD and correlation coefficients (MM/GBSA(+HB) model)}
\linespread{1.2}
\begin{table}[!h]
\begin{center}
\begin{tabular}[h]{ccccc}
\hline
\multicolumn{1}{c}{} & \multicolumn{1}{c}{} &  \multicolumn{2}{c}{\bf MM/GBSA+(HB)} & \\
\multicolumn{1}{c}{} & \multicolumn{1}{c}{} &  \multicolumn{1}{c}{\bf rmsd} & \multicolumn{1}{c}{\bf r} \\
\hline
\hline
\bf 0.0-0.5 ns &  & 0.993 & 0.534    \\
\bf 0.5-1.0 ns &  & 1.126 & 0.284   \\
\hline
\hline
\bf 0.0-1.0 ns &  & 1.098 & 0.356    \\
\hline
\end{tabular}
\caption{RMSD and correlation coefficients (r) between experimental and calculated values obtained averaging over different time intervals, using the MM/GBSA(+HB) model.}
\label{rmsd_corr_coeff}
\end{center}
\end{table}

\newpage

\subsection*{Table 7 - Optimal scaling factors (MM/GBSA(+HB) model)}
\linespread{0.9}
\begin{table}[!h]
\begin{center} 
\begin{tabular}[t]{ccccc}
\hline
\multicolumn{2}{c}{\bf MM/GBSA+(HB)}      &  \multicolumn{1}{c}{\bf 0.0-0.5 ns} & \multicolumn{1}{c}{\bf 0.5-1.0 ns} & \multicolumn{1}{c}{\bf 0.0-1.0 ns} \\
\hline
\multicolumn{2}{c}{ \bf $x_{Coul} $ } &  0.052 (0.028)  &  0.012 (0.024) &  0.039 (0.029)\\ 
\multicolumn{2}{c}{ \bf $x_{vdW}$ }   &  0.118 (0.049)  & -0.039 (0.055) &  0.060 (0.064)\\
\multicolumn{2}{c}{ \bf $x_{GB}$ }    &  0.054 (0.029)  &  0.014 (0.025) &  0.040 (0.031)\\
\multicolumn{2}{c}{ \bf $x_{H}$ }     & -0.001 (0.005)  &  0.007 (0.005) &  0.004 (0.006)\\
\multicolumn{2}{c}{ \bf $x_{P}$ }     & -0.013 (0.004)  &  0.001 (0.004) & -0.007 (0.005)\\
\multicolumn{2}{c}{ \bf $x_{HB}$ }    & -0.039 (0.120)  &  0.078 (0.107) &  0.027 (0.145)\\
\multicolumn{2}{c}{\bf $x_{const}$}   & -10.184 (6.002) &  3.055 (4.486) & -2.953 (6.527)\\
\hline
\end{tabular}
\caption{Optimal scaling factors for the MM/GBSA(+HB) model obtained averaging over different time intervals. Standard deviations (see Methods section) are given in parentheses.}
\label{Opt_gb_av}
\end{center} 
\end{table}

\newpage

\subsection*{Table 8 - Correct predictions}
\linespread{0.9}
\begin{table}[!h]
\begin{center}
{\scriptsize
\begin{tabular}[h]{cccccc}
\hline
 & $\Delta G_{calc}$ $<$ 1.0, & $\Delta G_{calc}$ $>$ 1.0, & $\Delta G_{calc}$ $\lessgtr$1.0,     &          &               \\
 & $\Delta G_{exp}$  $<$ 1.0 & $\Delta G_{exp}$ $>$ 1.0    & $\Delta G_{exp} $ $\gtrless$1.0,$D<0.5$ & \bf tot  &  ($D <$ 0.5)  \\   
\hline
\bf MM/DDDC-OONS (1r) &  10  & 23 & 2 & \bf 35 & (7) \\
\bf MM/DDDC-OONS (2r) & 15  & 24 & 0 & \bf 39 & (11) \\ 
\bf MM/DDDC-OONS (4r) & 16  & 22 & 1 & \bf 39 & (13) \\
\bf MM/DDDC-OONS (8r) & 14  & 23 & 1 & \bf 38 & (16) \\
\bf MM/DDDC-OONS(+HB) (1r) &  8  & 23 & 2 & \bf 33 & (9) \\
\bf MM/DDDC-OONS(+HB) (2r) & 14  & 23 & 1 & \bf 38 & (8) \\ 
\bf MM/DDDC-OONS(+HB) (4r) & 17  & 21 & 1 & \bf 39 & (10) \\
\bf MM/DDDC-OONS(+HB) (8r) & 15  & 22 & 2 & \bf 39 & (13) \\
\bf MM/DDDC-HP (1r) &  11  & 23 & 1 & \bf 35 & (12) \\
\bf MM/DDDC-HP (2r) & 16  & 21 & 4 & \bf 41 & (11) \\ 
\bf MM/DDDC-HP (4r) & 13  & 21 & 1 & \bf 35 & (12) \\
\bf MM/DDDC-HP (8r) & 13  & 24 & 0 & \bf 37 & (12) \\
\bf MM/DDDC-HP(+HB) (1r) &  12  & 23 & 1 & \bf 36 & (11) \\
\bf MM/DDDC-HP(+HB) (2r) & 15  & 22 & 3 & \bf 40 & (11) \\ 
\bf MM/DDDC-HP(+HB) (4r) & 13  & 21 & 2 & \bf 36 & (10) \\
\bf MM/DDDC-HP(+HB) (8r) & 16  & 24 & 1 & \bf 38 & (13) \\
\bf MM/GBSA & 13  & 24 & 1 & \bf 38 & (10) \\ 
\bf MM/GBSA(+HB) & 13  & 24 & 2 & \bf 39 & (13) \\ 
\hline
\end{tabular}}
\caption{Number of correct predictions of every model. Parenthetical data correspond to the number of prediction separated by less than
0.3 kcal/mol from the experimental data.}
\label{correct_preds}
\end{center}
\end{table}

\end{document}